\documentclass[twocolumn]{aastex61}
\usepackage{longtable}
\usepackage{graphicx}
\usepackage{soul}

\received{}
\revised{}
\accepted{}
\shorttitle{}
\shortauthors{Kosiarek et al. 2018}
\begin{document}
\newcommand{\kepler}{\textit{Kepler}}
\newcommand{\Spitzer}{\textit{Spitzer}}
\newcommand{\kt}{\textit{K2}}
\newcommand{\ktt}{K2-3}
\newcommand{\kttb}{K2-3 b}
\newcommand{\kttc}{K2-3 c}
\newcommand{\kttd}{K2-3 d}
\newcommand{\spitzer}{\textit{Spitzer}}
\newcommand\rotper{21.54 $\pm$ 0.49}
\newcommand{\gjmass}{12.58$^{+1.31}_{-1.28}$}
\newcommand{\kbmass}{6.48$^{+0.99}_{-0.93}$}
\newcommand{\kcmass}{2.14$^{+1.08}_{-1.04}$}
\newcommand{\kdmass}{2.80}
\newcommand{\mearth}{M$_\oplus$}
\newcommand{\rearth}{R$_\oplus$}
\newcommand{\krotper}{40 $\pm$ 2}
\newcommand{\ms}{m s$^{-1}$}

\title{Bright Opportunities for Atmospheric Characterization of Small Planets: Masses and Radii of K2-3 b, c, d and GJ3470 b from Radial Velocity Measurements and Spitzer Transits}

\correspondingauthor{Molly R. Kosiarek}
\email{mkosiare@ucsc.edu}

\author{Molly R. Kosiarek}
\affil{Department of Astronomy and Astrophysics, University of California, Santa Cruz, CA 95064, USA}
\affil{NSF Graduate Research Fellow}

\author{Ian J.M. Crossfield}
\affil{Department of Physics and Kavli Institute for Astrophysics and Space Research, Massachusetts Institute of Technology, Cambridge, MA 02139, USA}
\affil{Department of Astronomy and Astrophysics, University of California, Santa Cruz, CA 95064, USA}

\author{Kevin K. Hardegree-Ullman}
\affil{Department of Physics and Astronomy, The University of Toledo, Toledo, OH 43606}

\author{John H. Livingston}
\affil{Department of Astronomy, University of Tokyo, 7-3-1 Hongo, Bunkyo-ku, Tokyo 113-0033, Japan}
\affil{JSPS Fellow}

\author{Bj\"{o}rn Benneke}
\affil{D\'{e}partement de Physique, Universit\'{e} de Montr\'{e}al, 2900 Boulevard \'{E}‰douard-Montpetit, Montreal, Quebec H3T 1J4, Canada}

\author{Sarah Blunt}
\affil{California Institute of Technology, Pasadena, CA 91125, USA}

\author{Gregory W. Henry}
\affil{Center of Excellence in Information Systems, Tennessee State University, Nashville, TN  37209, USA}

\author{Ward S. Howard}
\affil{Department of Physics and Astronomy, University of North Carolina at Chapel Hill, Chapel Hill, NC 27599-3255, USA}

\author{David Berardo}
\affil{Department of Physics and Kavli Institute for Astrophysics and Space Research, Massachusetts Institute of Technology, Cambridge, MA 02139, USA}

%HIRES group
\author{Benjamin J. Fulton}
\affil{California Institute of Technology, Pasadena, CA 91125, USA}
\affil{Texaco Fellow}
\author{Lea A. Hirsch}
\affil{Astronomy Department, University of California, Berkeley, CA 94720, USA}
\author{Andrew W. Howard}
\affil{California Institute of Technology, Pasadena, CA 91125, USA}
\author{Howard Isaacson}
\affil{Astronomy Department, University of California, Berkeley, CA 94720, USA}
\author{Erik A.\ Petigura}
\affil{California Institute of Technology, Pasadena, CA 91125, USA}
\affil{Hubble Fellow}
\author{Evan Sinukoff}
\affil{Institute for Astronomy, University of Hawai`i at M\={a}noa, Honolulu, HI 96822, USA}
\affil{Cahill Center for Astrophysics, California Institute of Technology, 1216 East California Boulevard, Pasadena, CA 91125, USA}
\author{Lauren Weiss}
\affil{University of Montreal, Montreal, QC, H3T 1J4, Canada}
\affil{Trottier Fellow}

%K2 or SE data
\author{X.~Bonfils}
\affil{Univ. Grenoble Alpes, CNRS, IPAG, 38000 Grenoble, France}
\author{Courtney D. Dressing}
\affil{Astronomy Department, University of California, Berkeley, CA 94720, USA}
%and my ORCID ID is 0000-0001-8189-0233.
\author{Heather A. Knutson}
\affil{Division of Geological and Planetary Sciences, California Institute of Technology, Pasadena, CA 91125 USA}
\author{Joshua E. Schlieder}
\affil{Exoplanets and Stellar Astrophysics Laboratory, Code 667, NASA Goddard Space Flight Center, Greenbelt, MD, USA}

%Spitzer
\author{Michael Werner}
\affil{JPL/Caltech, Pasadena CA, 91107}
\author{Varoujan Gorjian}
\affil{JPL/Caltech, Pasadena CA, 91107}
\author{Jessica Krick}
\affil{Spitzer Science Center, Caltech, Pasadena CA 91125}
\author{Farisa Y. Morales}
\affil{Jet Propulsion Laboratory, California Institute of Technology, 4800 Oak Grove Drive, Pasadena, CA 91109, USA}
\affil{Moorpark College, 7075 Campus Rd, Moorpark, CA 93021, USA}
\affil{California State University Northridge, 18111 Nordhoff St, Northridge, CA 91330, USA}

%HARPS team
\author{Nicola Astudillo-Defru}
\affil{Observatoire de Gen\`eve, Universit\'e de Gen\`eve, 51 ch. des Maillettes, 1290 Sauverny, Switzerland}
\author{J.-M.~Almenara}
\affil{Observatoire de Gen\`eve, Universit\'e de Gen\`eve, 51 ch. des Maillettes, 1290 Sauverny, Switzerland}
\author{X.~Delfosse}
\affil{Univ. Grenoble Alpes, CNRS, IPAG, 38000 Grenoble, France}
\author{T.~Forveille}
\affil{Univ. Grenoble Alpes, CNRS, IPAG, 38000 Grenoble, France}
\author{C.~Lovis}
\affil{Observatoire de Gen\`eve, Universit\'e de Gen\`eve, 51 ch. des Maillettes, 1290 Sauverny, Switzerland}
\author{M.~Mayor}
\affil{Observatoire de Gen\`eve, Universit\'e de Gen\`eve, 51 ch. des Maillettes, 1290 Sauverny, Switzerland}
\author{F.~Murgas}
\affil{Instituto de Astrof\'sica de Canarias (IAC), E-38200 La Laguna, Tenerife, Spain}
\affil{Dept. Astrof\'isica, Universidad de La Laguna (ULL), E-38206 La Laguna, Tenerife, Spain}
\author{F.~Pepe}
\affil{Observatoire de Gen\`eve, Universit\'e de Gen\`eve, 51 ch. des Maillettes, 1290 Sauverny, Switzerland}
\author{N.~C.~Santos}
\affil{Instituto de Astrof\'isica e Ci\^encias do Espa\c{c}o, Universidade do Porto, CAUP, Rua das Estrelas, 4150-762 Porto, Portugal}
\affil{Departamento de F\'isica e Astronomia, Faculdade de Ci\^encias, Universidade do Porto, Rua do Campo Alegre, 4169-007 Porto, Portugal}
\author{S.~Udry}
\affil{Observatoire de Gen\`eve, Universit\'e de Gen\`eve, 51 ch. des Maillettes, 1290 Sauverny, Switzerland}

%Everyscope group
\author{H. T. Corbett}
\affil{Department of Physics and Astronomy, University of North Carolina at Chapel Hill, Chapel Hill, NC 27599-3255, USA}
\author{Octavi Fors}
\affil{Department of Physics and Astronomy, University of North Carolina at Chapel Hill, Chapel Hill, NC 27599-3255, USA}
\affil{Institut de Ciencies del Cosmos (ICCUB), Universitat de Barcelona, IEEC-UB, Marti i Franques 1, E08028 Barcelona, Spain}
\author{Nicholas M. Law}
\affil{Department of Physics and Astronomy, University of North Carolina at Chapel Hill, Chapel Hill, NC 27599-3255, USA}
\author{Jeffrey K. Ratzloff}
\affil{Department of Physics and Astronomy, University of North Carolina at Chapel Hill, Chapel Hill, NC 27599-3255, USA}
\author{Daniel del Ser}
\affil{Department of Physics and Astronomy, University of North Carolina at Chapel Hill, Chapel Hill, NC 27599-3255, USA}
\affil{Dept. Fisica Quantica i Astrofisica (FQA). Institut de Ciencies del Cosmos (ICCUB), Universitat de Barcelona, UB, Marti i Franques 1, E08028 Barcelona, Spain}

\begin{abstract}
We report improved masses, radii, and densities for four planets in two bright M-dwarf systems, \ktt\ and GJ3470, derived from a combination of new radial velocity and transit observations. Supplementing \kt\ photometry with follow-up \spitzer\ transit observations refined the transit ephemerides of K2-3 b, c, and d by over a factor of 10.  We analyze ground-based photometry from the Evryscope and Fairborn Observatory to determine the characteristic stellar activity timescales for our Gaussian Process fit, including the stellar rotation period and activity region decay timescale. The stellar rotation signals for both stars are evident in the radial velocity data and are included in our fit using a Gaussian process trained on the photometry. We find the masses of K2-3 b, K2-3 c and GJ3470 b to be \kbmass, \kcmass, and \gjmass\ \mearth\ respectively. K2-3 d was not significantly detected and has a 3$\sigma$ upper limit of \kdmass\ \mearth. These two systems are training cases for future TESS systems; due to the low planet densities ($\rho$ $<$ 3.7 g cm$^{-3}$) and bright host stars (K $<$ 9 mag), they are among the best candidates for transmission spectroscopy in order to characterize the atmospheric compositions of small planets. 
\end{abstract}

\keywords{techniques: radial velocities, techniques: photometric, planets and satellites: composition, }

\section{Introduction}

The field of exoplanets has shifted from detection to characterization due to technological improvements in instrumentation and large detection surveys such as NASA's \kepler\ mission. One of the most surprising results from \kepler\ was the prevalence of planets between 1 and 4R$_\oplus$, called super-Earths€ or sub-Neptunes€, which are absent from our solar system \citep{Howard2012,Dressing2013,Dressing2015,Fressin2013,Petigura2013}. Planets of this size occur more frequently around M stars than G or F stars (\citet{Mulders2015} for orbital periods of $<$ 150 days). 

Core-accretion models predict that an intermediate sized planet will become the core of a gas giant through runaway gas accretion. Therefore, these models are at odds with the prevalence of such intermediate mass planets \citep{Mizuno1980,Bodenheimer2014}. To avoid this problem, \citet{Lee2014} and \citet{Lee2016} proposed that super-Earths formed later than gas giants, without time to undergo runaway gas accretion. It is also debated whether there is sufficient material in the inner protoplanetary disk to form these planets \citep{Weidenschilling1977,Hayashi1981}. Pebble accretion and migration could address this problem and form closely packed multiplanet systems orbiting M dwarfs \citep{Swift2013,Ormel2017}. 

Planet compositions provide a crucial link to their formation histories. The composition can be inferred either from the bulk density, which is derived from the planet's mass and radius, or from atmospheric studies. \kepler\ transits and ground-based radial velocity (RV) follow-up discovered an increase in bulk density with decreasing size, suggesting a transition region at 1.5--2.0 R$_{\oplus}$ between volatile-rich gas/ice planets and rocky planets \citep{Weiss2014,Rogers2015,Fulton2017,Eylen2017}.

Due to the approaching launch of the James Webb Space Telescope (JWST) and selection for future European Space Agency (ESA) mission ARIEL, preparatory measurements of potential atmospheric characterization targets are important for identifying the best targets as well as for the interpretation of the spectra. Primarily, target ephemerides must be refined in order to reduce the transit timing uncertainty and therefore use space-based time most efficiently. Furthermore, precise mass measurements and surface gravity calculations are necessary as these parameters will affect the interpretation of the transmission spectra. Both atmospheric scale height and molecular absorption affect the depth of the planet's spectroscopic features. Since atmospheric scale height is related to the surface gravity, a precise mass measurement is needed in order to correctly interpret the molecular absorption features in a spectrum \citep{Batalha2017}.

The \kt\ mission has discovered many cool planets orbiting bright stars
\citep{Montet2015,Crossfield2016,Vanderburg2016,Dressing2017b,Mayo2018},
and TESS will find a large sample of even brighter systems around nearby stars \citep{Ricker2014,Sullivan2015}. These bright host stars can be more precisely followed up from ground-based telescopes and are amenable to transmission spectroscopy observations. This paper illustrates a follow-up program to prepare for potential JWST observations of two systems much like those that will be found by TESS.

In this paper we describe precise RV and photometry follow-up of two systems, K2-3 and GJ3470. Both of these systems have sub-Neptune-sized planets orbiting M-dwarf stars and are amenable to atmospheric transmission spectroscopy. In Section 2 we describe the two systems. In Section 3 we detail our \spitzer\ observations and analysis. In Section 4 we describe our RV analysis and related photometric follow-up, then present our RV results. In Section 5 we examine these planets in the context of other similar sub-Neptune systems and discuss atmospheric transmission spectroscopy considerations before concluding in Section 6.

\section{Target Systems and Stellar Parameters}

K2-3 (EPIC 201367065) is a bright (Ks = 8.6 mag), nearby (45 $\pm$ 3 pc) M0 dwarf star hosting three planets from 1.5 to 2 R$_{\oplus}$ at orbital periods between 10 and 45 days \citep{Crossfield2015}(\autoref{tab:K23stellarprop}). These planets receive 1.5--10 times the flux incident on Earth; planet d orbits near the habitable zone. 

\begin{deluxetable}{lccc}
\tablecaption{K2-3 Stellar Properties \label{tab:K23stellarprop}}
\tablehead{\colhead{Parameter} & \colhead{Value} & \colhead{Units} & \colhead{Source}}
\startdata
\sidehead{\bf{Identifying Information}}
RA & 11:29:20.388 & & (1) \\
DEC & -01:27:17.23 & & (1) \\
\sidehead{\bf{Photometric Properties}}
\textit{J} & 9.421 $\pm$ 0.027 & mag & (2) \\
\textit{H} & 8.805 $\pm$ 0.044 & mag & (2) \\
\textit{K} & 8.561 $\pm$ 0.023 & mag & (2) \\
\textit{Kp} & 11.574 & mag & (3) \\
Rotation Period & \krotper & days & (4) \\
\sidehead{\bf{Spectroscopic Properties}}
Barycentric RV & 32.6 $\pm$ 1 & km s$^{-1}$ & (1) \\
Distance & 45 $\pm$ 3 & pc & (1) \\
H$\alpha$ & 0.38 $\pm$ 0.06 & Ang & (1) \\
Age  & $\ge1$ & Gyr & (1) \\
Spectral Type & M0.0 $\pm$ 0.5 V & & (1) \\
$[$Fe/H$]$ & -0.32 $\pm$ 0.13 & & (1) \\
Temperature  & 3896 $\pm$ 189 & K & (1) \\
Mass & 0.601 $\pm$ 0.089 & M$_{\odot}$ & (1) \\
Radius  & 0.561 $\pm$ 0.068 & R$_{\odot}$ & (1) \\
Density & 3.58 $\pm$ 0.61 & $\rho_{\odot}$& (5) \\
Surface Gravity  & 4.734 $\pm$ 0.062 & cgs & (5) \\
\enddata
\tablecomments{(1) \citet{Crossfield2015}, (2) \citet{Cutri2003}, (3) \citet{Huber2016}, (4) this work, (5) \citet{Almenara2015}. }
\end{deluxetable}

K2-3 b, c, and d were discovered in \kt\ photometry \citep{Crossfield2015}. Since then, there have been multiple RV and transit follow-up measurements. \citet{Almenara2015} collected 66 HARPS spectra and determined the masses of planet b, c, and d to be $8.4 \pm 2.1$, $2.1^{+2.1}_{-1.3}$, and $11.1 \pm 3.5$ \mearth, respectively. \citet{Almenara2015} caution that the RV semi-amplitudes of planets c and d are likely affected by stellar activity. 
\citet{Dai2016} collected 31 spectra with the Planet Finder Spectrograph (PFS) on Magellan and modeled the RV data with Almenara's HARPS data. The combined datasets constrained the masses of planets b, c, and d to be $7.7 \pm 2.0$, $<12.6$, and $11.3^{+5.9}_{-5.8}$ \mearth, respectively. 
\citet{Damasso2018} performed an RV analysis on a total of 132 HARPS spectra and 197 HARPS-N spectra, including the Almenara sample. This HARPS analysis found the mass of planet b and c to be 6.6 $\pm$ 1.1 and 3.1$^{+1.3}_{-1.2}$ \mearth, respectively. The mass of planet d is estimated as 2.7$^{+1.2}_{-0.8}$ from a suite of injection-recovery tests. 
\citet{Beichman2016} refined the ephemeris and radii of the three planets with seven follow-up \spitzer\ transits and \citet{Fukui2016} observed a ground-based transit of K2-3 d to further refine its ephemeris. 

GJ3470 is also a bright ($K$ = 8.0 mag), nearby (29.9$^{+3.7}_{-3.4}$ pc) M1.5 dwarf hosting one Neptune-sized planet in a 3.33 day orbit \citep{Cutri2003,Bonfils2012}(\autoref{tab:GJstellarprop}). GJ3470 b was discovered in a HARPS RV campaign that searched for short-period planets orbiting M dwarfs and was subsequently observed in transit. GJ3470 b has an equilibrium temperature near 700 K and a radius of 3.9 R$_{\oplus}$. Its mass has been measured previously to be 13.73$\pm$1.61, 14.0$\pm$1.8, and 13.9$^{+1.5}_{-1.4}$ \mearth\ by \citet{Bonfils2012}, \citet{Demory2013}, and \citet{Biddle2014} respectively. 
Its low density supports a substantial atmosphere covering the planet \citep{Biddle2014}. Seven previous studies have investigated its atmospheric composition. \citet{Fukui2013} found variations in the transit depths in the $J$, $I$, and 4.5 $\mu$m bands that suggest that the atmospheric opacity varies with wavelength due to the absorption or scattering of stellar light by atmospheric molecules. \cite{Nascimbeni2013} detected a transit depth difference between the ultraviolet and optical wavelengths also indicating a Rayleigh-scattering slope, confirmed by \citet{Biddle2014,Chen2017} and \citet{Dragomir2015}. \citet{Crossfield2013} found a flat transmission spectrum in the $K$-band suggesting a hazy, methane-poor, or high-metallicity atmosphere. Finally, \citet{Bourrier2018} find the planet is surrounded by a large exosphere of neutral hydrogen from Hubble Lyman alpha measurements.

\begin{deluxetable}{lccc}
\tablecaption{GJ3470 Stellar and Planet b Transit Properties \label{tab:GJstellarprop}}
\tablehead{\colhead{Parameter} & \colhead{Value} & \colhead{Units} & \colhead{Source}}
\startdata
\sidehead{\bf{Photometric Properties}}
Spectral type & M1.5 & & (1) \\
$V$ & 12.3 & mag & (2)\\
$K$ & 7.989 $\pm$ 0.023 & mag & (3) \\
$J$ & 8.794 $\pm$ 0.019 & mag & (3) \\
$H$ & 8.206 $\pm$ 0.023 & mag & (3) \\
Rotation Period & \rotper & days & (5) \\
\sidehead{\bf{Spectroscopic Properties}}
Luminosity & 0.029 ± 0.002 & L$_{\odot}$ & (2) \\ 
Mass  & 0.51 $\pm$ 0.06 & M$_{\odot}$ & (4) \\
Radius  & 0.48 $\pm$ 0.04 & R$_{\odot}$ & (4) \\
Distance & 30.7$^{+2.1}_{-1.7}$ & pc & (6) \\
Age & 0.3--3 & Gyr & (2)  \\
Temperature & 3652 $\pm$ 50 & K & (4) \\
Surface Gravity & 4.658 $\pm$ 0.035 & cgs & (6)\\
$[$Fe/H$]$ & +0.20 $\pm$ 0.10 & & (6) \\
\sidehead{\bf{Transit Properties}}
T$_0$ ($-2450000$) & 6677.727712 $\pm$ 0.00022 & BJD & (7)\\
T$_{14}$ & 0.07992$^{+0.00100}_{-0.00099}$ & days & (7) \\
P & 3.3366413$\pm$0.0000060 & days & (7) \\
R$_p$  & 3.88 $\pm$ 0.32 & \rearth &  (4) \\
a/R$_*$ & 12.92$^{+0.72}_{-0.65}$ & & (7)\\
T$_{eq}$ & 615 $\pm$ 16 & K & (2) \\
\enddata
\tablecomments{(1) \citet{Reid1997}, (2) \citet{Bonfils2012}, (3) \citet{Cutri2003}, (4) \citet{Biddle2014}, (5) this work, (6) \citet{Demory2013}, (7) \citet{Dragomir2015}. }
\end{deluxetable}

\section{K2-3 \spitzer\ Observations}
\label{sec:spitzer}

We observed six transits of \kttb, two transits of \kttc, and two transits of \kttd~using Channel 2 (4.5 $\mu$m) of the Infrared Array Camera (IRAC) on the \textit{Spitzer Space Telescope} to refine the transit parameters of these three planets (GO 11026, PI: Werner; GO 12081, PI: Benneke). K2-3 was observed in staring mode and placed on the ``sweet spot'' pixel to keep the star in one location during the observations and minimize the effect of gain variations. To minimize data volume and overhead from readout time, the subarray mode was used with an exposure time of 2 seconds per frame. This produced between 11392 and 26368 individual frames per observation. Total observation durations were between 6.5 and 15 hours to include adequate out-of-transit baseline and were typically centered near the predicted mid-transit time from the \kt\ ephemeris. 

In the following subsection, we describe two different analyses performed on these \spitzer\ data. The first analysis performs an individual fit to each \spitzer\ transit separately to check for consistency of the parameters between individual transit events whereas the second analysis performs a combined fit to the \spitzer\ data to derive global parameters.

\subsection{\spitzer\ Transit Analysis}

We extract the \spitzer\ light curves following the approach taken by \citet{2012ApJ...754...22K} and \citet{2016ApJ...822...39B}, using a circular aperture 2.4 pixels in radius centered on the host star. We used the Python package \texttt{photutils} \citep{Bradley2016} for centroiding and aperture photometry. 
We use a modified version of the pixel-level decorrelation (PLD, \citet{Deming2015}) adapted from \citet{Benneke2017} to simultaneously model the \Spitzer\ systematics (intra-pixel sensitivity variations) and the exoplanet system parameters. The instrument sensitivity is modeled by equation 1 in \citet{Benneke2017}, and we use the Python package \texttt{batman} \citep{Kreidberg2015} to generate the transit models. 
For parameter estimation we use the Python package \texttt{emcee} \citep{emcee}, an implementation of the affine-invariant Markov chain Monte Carlo (MCMC) ensemble sampler \citep{2010CAMCS...5...65G}. 
We find that using a $3 \times 3$ pixel grid sufficiently captures the information content corresponding to the motion of the point-spread function (PSF) on the detector (which is typically $\lesssim$ a few tenths of a pixel). %\autoref{fig:pixels} shows the light curve on each pixel for the observed transits. 
In comparisons between various methods used to correct \spitzer\ systematics \citep{2016AJ....152...44I}, PLD was among the top performers, displaying both high precision and repeatability. For more details about this type of IRAC photometry analysis, see \citet{Livingston2019} and K. Hardegree-Ullman (2019, in preparation).

\begin{figure}[h]
\centering
\includegraphics[width=3.4in]{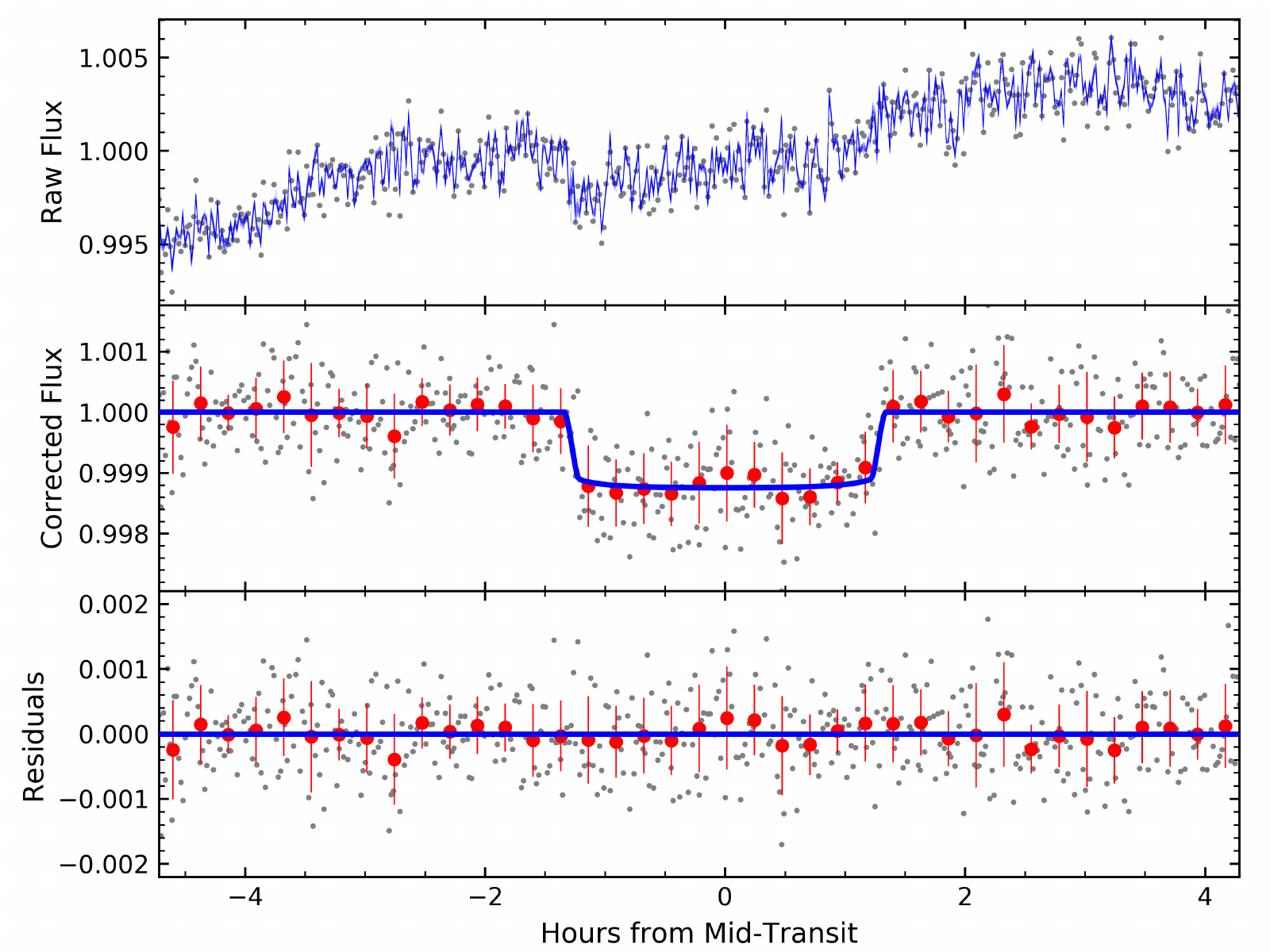}
\caption{Example fit to one transit of K2-3 b. Top: raw \spitzer\ data (black points) including the fit to the systematics (blue line). middle: Transit of K2-3 b including raw data (grey points), binned data (red points), and transit model (blue line). Bottom: residuals from the model. 
}
\label{fig:onetran}
\end{figure}

We first analyzed the \spitzer\ transits one at a time to check for consistency of parameters between independent transit events. 
For the individual transit models, we fit for the scaled planet radius $R_p/R_{\star}$, the mid-transit time $T_0$, the scaled semi-major axis $a/R_{\star}$, and the orbital inclination angle $i$. The quadratic limb-darkening coefficients for the 4.5$\mu$m \Spitzer\ bandpass were found by interpolating the values from \citet{Claret2011}. We held the orbital periods constant at the values found by \citet{Beichman2016}, and fixed eccentricity and longitude of periastron to 0, but note that these parameters have a negligible effect on the overall shape of the transit model. Gaussian priors were imposed on the transit system parameters based on our previous knowledge of the system from \citet{2015ApJ...804...10C}. We also found global system parameters by combining the posterior distributions from each individual result for each planet and finding the 16th, 50th, and 84th percentiles of the combined distribution. 
One example transit is shown in~\autoref{fig:onetran}.

\begin{figure}[h]
\hspace*{-0.5cm}  
\includegraphics[width=3.9in]{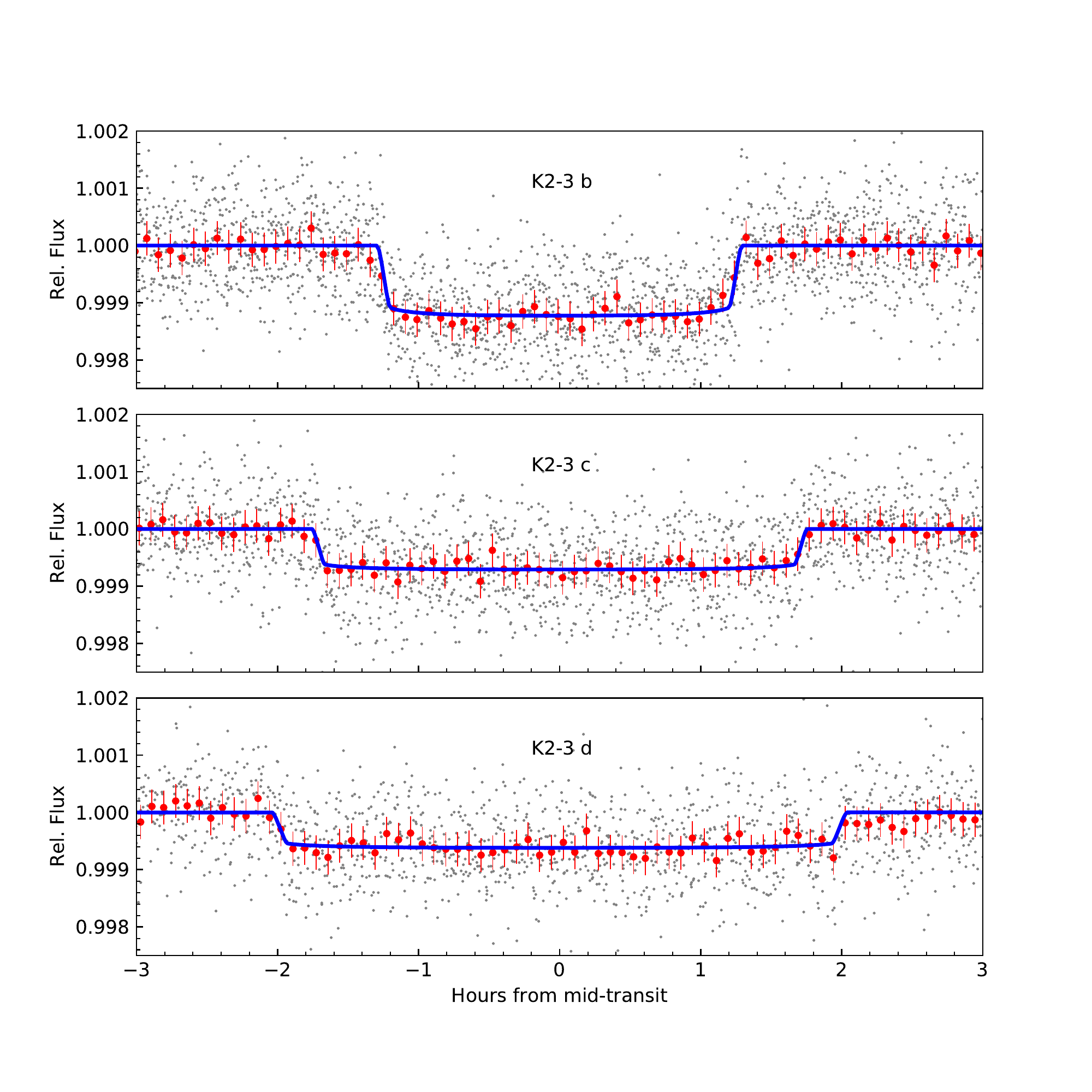}
\caption{Transit fits to K2-3 b, c, and d displaying all of our \spitzer\ data for each planet. Individual \spitzer\ datapoints (grey), binned points (red), and planet fits from our simultaneous analysis (blue) for each K2-3 planet are shown. \label{fig:allspitzer}
}
\end{figure}

We then perform our global fit to the 10 \spitzer\ transit datasets of the three planets in this system by constructing a joint model comprising a set of shared transit parameters and a set of systematic parameters corresponding to each individual transit observation. We adopt a global quadratic limb-darkening law and set Gaussian priors on the parameters $u_1$ and $u_2$ by interpolating the table of \citet{2012yCat..35460014C}, where the widths are determined by a Monte Carlo simulation. To restrict exploration of limb-darkening parameter space to only physical scenarios, we utilize the triangular sampling method of \citet{2013MNRAS.435.2152K}; thus, we actually sample in $q_1$/$q_2$ space. For each of the three planets in the system, we use a unique set of transit parameters: the period $P$, time of mid-transit $T_0$, planet-star radius ratio $R_P/R_{\star}$, scaled semi-major axis $a/R_{\star}$, and impact parameter $b$. The rest of the parameters in the model correspond to the PLD coefficients for each individual dataset. Besides the Gaussian priors on the limb-darkening parameters, we also impose Gaussian priors on $T_0$, $P$, and the mean stellar density of the host star, based on the values reported in \citet{2015ApJ...804...10C}. See~\autoref{fig:allspitzer} for the transit fits for K2-3 b, c, and d obtained from our global \spitzer\ analysis. 

All of our transit parameters are shown in~\autoref{tab:spitzer} along with the parameters derived from only the \kt\ transits for comparison \citep{Crossfield2015}. We combined the parameters from the individual fits by adding their posteriors in order to compare the individual fits with the global analysis. The parameters from these two analyses are all within 1$\sigma$. We adopt the parameters from the simultaneous transit analysis for our RV analysis.

\begin{deluxetable*}{ccccccccc}
\tablecaption{K2-3 \spitzer\ Transit Fit Parameters \label{tab:spitzer}}
\tablehead{\colhead{Planet} & \colhead{P [days]} & \colhead{Mid-Transit [BJD]} & \colhead{$R_{p}/R_{\star}$ [\%]} & \colhead{$R_{p}/R_{\oplus}$} & \colhead{$a/R_{\star}$} & \colhead{$i$ [$^{\circ}$]} & \colhead{$b$} & \colhead{Source}} 
\startdata
b & -- & $2457094.94680^{+0.00080}_{-0.00068}$ & $3.494^{+0.087}_{-0.087}$ & $2.134^{+0.272}_{-0.260}$ & $29.44^{+3.89}_{-3.91}$ & $89.50^{+0.35}_{-0.43}$ & $0.26^{+0.21}_{-0.17}$ & IRAC2 Transit 1 \\
b & -- & $2457105.00241^{+0.00068}_{-0.00062}$ & $3.864^{+0.094}_{-0.095}$ & $2.366^{+0.298}_{-0.300}$ & $29.478^{+3.93}_{-3.88}$ & $89.56^{+0.30}_{-0.39}$ & $0.23^{+0.19}_{-0.16}$ & IRAC2 Transit 2 \\
b & -- & $2457275.92778^{+0.00149}_{-0.00121}$ & $3.355^{+0.090}_{-0.090}$ & $2.053^{+0.259}_{-0.258}$ & $29.54^{+3.91}_{-3.99}$ & $89.31^{+0.45}_{-0.50}$ & $0.35^{+0.25}_{-0.23}$ & IRAC2 Transit 3 \\
b & -- & $2457466.97119^{+0.00114}_{-0.00086}$ & $3.360^{+0.084}_{-0.085}$ & $2.057^{+0.253}_{-0.260}$ & $29.43^{+3.89}_{-3.95}$ & $89.46^{+0.37}_{-0.46}$ & $0.27^{+0.23}_{-0.18}$ & IRAC2 Transit 4 \\
b & -- & $2457497.13286^{+0.00107}_{-0.00111}$ & $3.659^{+0.088}_{-0.088}$ & $2.237^{+0.286}_{-0.271}$ & $29.52^{+4.02}_{-4.00}$ & $89.42^{+0.40}_{-0.49}$ & $0.30^{+0.24}_{-0.20}$ & IRAC2 Transit 5 \\
b & -- & $2457627.84302^{+0.00090}_{-0.00232}$ & $3.572^{+0.090}_{-0.091}$ & $2.192^{+0.267}_{-0.266}$ & $29.46^{+3.89}_{-3.92}$ & $89.52^{+0.33}_{-0.44}$ & $0.25^{+0.21}_{-0.17}$ & IRAC2 Transit 6 \\
b & $10.054638\pm0.000016$ & -- & $3.532^{+0.243}_{-0.185}$ & $2.165^{+0.297}_{-0.284}$ & $29.43^{+3.93}_{-3.91}$ & $89.47^{+0.37}_{-0.46}$ & $0.27^{+0.22}_{-0.19}$ & Individual Combined \\
b & $10.054626^{+0.000009}_{-0.000010}$ & $2456813.41843^{+0.00039}_{-0.00038}$ &$3.44^{+0.04}_{-0.04}$   & $2.103^{+0.257}_{-0.256}$  & $30.02^{+0.25}_{-0.31}$  & $89.588^{+0.116}_{-0.100}$ & $0.22^{+0.05}_{-0.06}$ & Simultaneous Fit \\
b & 10.05403$^{+0.00026}_{-0.00025}$ & 2456813$\pm$0.0011 & 3.483$^{+0.123}_{-0.070}$ & 2.14$^{+0.27}_{-0.26}$ & 29.2$^{+1.8}_{-3.6}$ & 89.28$^{+0.46}_{-0.60}$ & 0.37$^{+0.22}_{-0.23}$ & \citet{Crossfield2015} \\
\hline
c & -- & $2457108.03664^{+0.00172}_{-0.00187}$ & $2.549^{+0.080}_{-0.081}$ & $1.557^{+0.197}_{-0.190}$ & $53.62^{+7.12}_{-7.13}$ & $89.73^{+0.21}_{-0.27}$ & $0.29^{+0.24}_{-0.19}$ & IRAC2 Transit 1 \\
c & -- & $2457280.56131^{+0.00215}_{-0.00224}$ & $2.554^{+0.080}_{-0.081}$ & $1.559^{+0.198}_{-0.195}$ & $53.28^{+7.19}_{-7.23}$ & $89.69^{+0.21}_{-0.27}$ & $0.28^{+0.24}_{-0.20}$ & IRAC2 Transit 2 \\
c & -- & $2457477.73145^{+0.00254}_{-0.00253}$ & $2.670^{+0.081}_{-0.081}$ & $1.635^{+0.204}_{-0.200}$ & $53.30^{+7.15}_{-7.07}$ & $89.72^{+0.19}_{-0.26}$ & $0.26^{+0.24}_{-0.18}$ & IRAC2 Transit 3 \\
c & -- & $2457625.61918^{+0.00171}_{-0.00199}$ & $2.726^{+0.077}_{-0.081}$ & $1.668^{+0.211}_{-0.207}$ & $53.49^{+7.10}_{-7.03}$ & $89.73^{+0.19}_{-0.25}$ & $0.26^{+0.22}_{-0.19}$ & IRAC2 Transit 4 \\
c & -- & $2457650.26528^{+0.00191}_{-0.00129}$ & $2.743^{+0.079}_{-0.079}$ & $1.680^{+0.209}_{-0.207}$ & $53.67^{+7.00}_{-7.01}$ & $89.72^{+0.19}_{-0.24}$ & $0.26^{+0.22}_{-0.18}$ & IRAC2 Transit 5 \\
c & $24.646569\pm0.000047$ & -- & $2.653^{+0.116}_{-0.128}$ & $1.618^{+0.212}_{-0.207}$ & $53.47^{+7.10}_{-7.15}$ & $89.71^{+0.20}_{-0.26}$ & $0.27^{+0.23}_{-0.19}$ & Individual Combined \\
c &$24.646582^{+0.000039}_{-0.000039}$  &$2456812.28013^{+0.00090}_{-0.00095}$ & $2.59^{+0.06}_{-0.06}$ & $1.584^{+0.197}_{-0.195}$& $54.57^{+0.46}_{-0.56}$ & $89.905^{+0.066}_{-0.088}$&$0.09^{+0.08}_{-0.06}$  & Simultaneous Fit \\
c & 24.6454$\pm$0.0013 & 2456812$^{+0.00026}_{-0.00025}$ & 2.786$^{+0.143}_{-0.083}$ & 1.72$^{+0.23}_{-0.22}$ & 51.8$^{+4.1}_{-9.1}$& 89.55$^{+0.29}_{-0.44}$ & 0.41$^{+0.26}_{-0.25}$ & \citet{Crossfield2015} \\
\hline
d & -- & $2457093.56831^{+0.00517}_{-0.00325}$ & $2.479^{+0.089}_{-0.089}$ & $1.521^{+0.191}_{-0.190}$ & $79.09^{+10.55}_{-10.73}$ & $89.81^{+0.13}_{-0.17}$ & $0.27^{+0.23}_{-0.19}$ & IRAC2 Transit 1 \\
d & -- & $2457271.79827^{+0.00477}_{-0.00359}$ & $2.490^{+0.088}_{-0.089}$ & $1.521^{+0.194}_{-0.191}$ & $79.74^{+10.63}_{-10.82}$ & $89.81^{+0.13}_{-0.17}$ & $0.27^{+0.23}_{-0.19}$ & IRAC2 Transit 2 \\
d & -- & $2457494.57861^{+0.00405}_{-0.00296}$ & $2.454^{+0.082}_{-0.083}$ & $1.503^{+0.184}_{-0.191}$ & $79.51^{+10.55}_{-10.67}$ & $89.79^{+0.14}_{-0.17}$ & $0.30^{+0.22}_{-0.20}$ & IRAC2 Transit 3 \\
d & -- & $2457628.23815^{+0.00891}_{-0.00200}$ & $2.449^{+0.081}_{-0.082}$ & $1.500^{+0.189}_{-0.185}$ & $79.69^{+10.78}_{-10.63}$ & $89.80^{+0.14}_{-0.17}$ & $0.29^{+0.24}_{-0.20}$ & IRAC2 Transit 4 \\
d & $44.556913\pm0.000182$ & --  & $2.468^{+0.086}_{-0.087}$ & $1.511^{+0.191}_{-0.193}$ & $79.35^{+10.71}_{-10.53}$ & $89.80^{+0.14}_{-0.17}$ & $0.28^{+0.23}_{-0.19}$ & Individual Combined \\
d &$44.556456^{+0.000097}_{-0.000087}$ & $2456826.22347^{+0.00053}_{-0.00052}$& $2.44^{+0.08}_{-0.08}$& $1.492^{+0.189}_{-0.186}$&$80.98^{+0.68}_{-0.84}$ & $89.788^{+0.033}_{-0.029}$& $0.30^{+0.04}_{-0.05}$& Simultaneous Fit \\
d & 44.5631$^{+0.0063}_{-0.0043}$ & 2456826$^{+0.00037}_{-0.00043}$ & 2.48$^{+0.14}_{-0.10}$ & 1.52$^{+0.21}_{-0.20}$ & 78.7$^{+6.7}_{-13}$& 89.68$^{+0.21}_{-0.26}$ & 0.45$^{+0.23}_{-0.28}$ & \citet{Crossfield2015} \\
\enddata
\end{deluxetable*}
\subsection{\spitzer\ Ephemeris Improvement}

\begin{figure}[h]
\hspace*{-0.5cm}  
\includegraphics[width=3.7in]{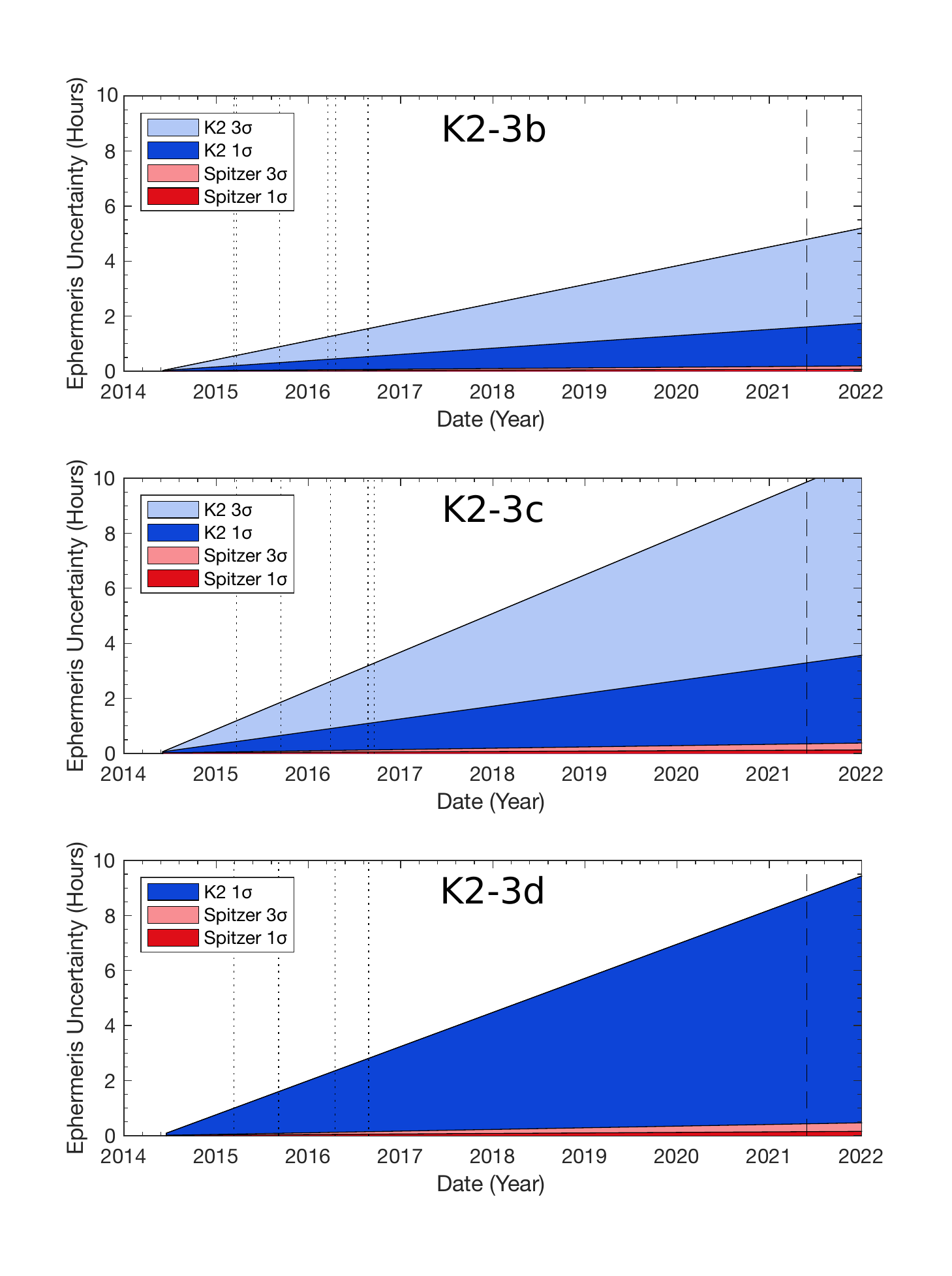}
\caption{Transit time uncertainty for the K2-3 system in the JWST era. The blue region illustrates the 3$\sigma$ uncertainty (light blue) and 1$\sigma$ uncertainty (dark blue) on the transit times from the K2-derived ephemerides reported by \citet{Crossfield2015}. Equivalently, the red region illustrates the 1$\sigma$ and 3$\sigma$ uncertainty on the transit times from our \spitzer-derived ephemerides. The vertical dotted lines illustrate the times of the \spitzer\ observations. The vertical dashed line shows the scheduled JWST launch in March 2021. The \spitzer\ transits decrease the uncertainty on the transit times by over a factor of 10. \label{fig:spitzertime}
}
\end{figure}

These \spitzer\ data reduce the uncertainty on the transit times and periods of the K2-3 planets. Refining the ephemerides is particularly important in order to efficiently subdivide time on large telescopes and space-based telescopes. ~\autoref{fig:spitzertime} shows the uncertainty of the transit time of each planet propagated forwards to 2022, shortly after the launch of JWST. This refinement is crucial to accurately schedule transit observations with future space-based atmospheric missions.
For example, if one wanted to observe K2-3 d in the JWST era, the \kt\ 3$\sigma$ uncertainty of the transit mid-point is over 25 hours; this would waste considerable telescope time and will only increase as the time baseline lengthens. \citet{Beichman2016} refined its ephemeris with \spitzer\ measurements two years ago. Our measurements further refine the orbital period uncertainty by a over factor of twenty from the original \kt\ data and to one-third of that from \citet{Beichman2016}. From our \spitzer\ analysis, the 3$\sigma$ uncertainty in the transit mid-point of K2-3 d in 2022 has improved to only 30 minutes.

\section{Radial Velocity Analysis}

\subsection{RV Observations}

We obtained RV measurements of K2-3 and GJ3470 using the High Resolution Echelle Spectrometer (HIRES)\citep{Vogt1994} on the Keck I Telescope. We collected 74 measurements of K2-3 from 2015 Feb 4 to 2017 Apr 11 and 56 measurements of GJ3470 from 2012 Sep 25 to 2017 Mar 15. These spectra were taken with an iodine cell and the C2 decker; a template spectrum was also taken in order to calibrate the wavelength and estimate the RV uncertainty. On average, measurements of K2-3 were collected with an exposure time of 1600s in order to reach a signal-to-noise ratio (S/N) of 87/pixel (80k counts on the HIRES exposure meter). Measurements of GJ3470 were collected with an exposure time of 1200s in order to reach a S/R of 60/pixel (40k counts). The observations and data reduction followed the California Planet Search method described in \citet{Howard2010}. RV data for K2-3 and GJ3470 are shown in~\autoref{tab:k23rv} and~\autoref{tab:gj3470rv}. 

\startlongtable
\begin{deluxetable}{cccc}
\tablecaption{K2-3 HIRES Relative RV Measurements  \label{tab:k23rv}}
\setlength\extrarowheight{-1.5pt}
\tablehead{\colhead{BJD$_{TDB}$}& \colhead{RV (m/s)} & \colhead{Unc. (m/s)} & \colhead{$S_\mathrm{HK}$ $\pm$ 0.005} }
\startdata
2457057.93921&0.366715&2.333335&0.9000  \\
2457058.03821&-0.957106&1.734197&0.7805  \\
2457058.05976&-3.153888&1.63551&0.7963  \\
2457058.08085&-7.341192&1.583017&0.7784  \\
2457058.92564&11.883496&1.898559&0.7086  \\
2457058.95076&10.980076&2.34516&0.6743  \\
2457058.97218&3.558983&2.28085&0.7210  \\
2457059.08507&2.8289&1.947861&0.6640  \\
2457059.10630&0.20892&1.917224&0.5664  \\
2457059.13391&-1.2014&1.994416&0.5140  \\
2457061.99505&0.377411&1.736609&0.7832  \\
2457062.00959&3.934113&1.810443&0.8032  \\
2457062.02387&4.739629&1.640702&0.8091  \\
2457062.09685&-3.780676&1.574131&0.7972  \\
2457062.11334&2.534422&1.596577&0.8000  \\
2457062.13098&4.418034&1.67832&0.7943  \\
2457150.95144&-8.490637&1.916942&0.7797  \\
2457150.96587&-7.028363&1.936422&0.7828  \\
2457179.79623&-8.162854&1.559401&0.7842  \\
2457179.81064&-8.641193&1.509294&0.7974  \\
2457179.82504&-10.420957&1.724975&0.7913  \\
2457200.80146&-1.273851&1.799606&0.8192  \\
2457200.81577&-2.722842&1.907786&0.7689  \\
2457201.83038&-4.710788&2.031547&0.7397  \\
2457208.77055&-6.80142&1.712294&0.7549  \\
2457210.79928&-4.84792&2.040615&0.7334  \\
2457213.81025&7.015119&1.824659&0.5133  \\
2457216.77797&-8.124406&2.05162&0.6954  \\
2457353.10152&4.475981&1.543244&0.7053  \\
2457353.12056&0.028874&1.546589&0.7198  \\
2457353.13925&-0.068528&1.601731&0.7463  \\
2457354.10453&-0.48829&1.809439&0.7089  \\
2457355.10054&1.319467&1.438562&0.7483  \\
2457355.12162&-1.344597&1.361441&0.7516  \\
2457355.14113&-1.368156&1.456596&0.7424  \\
2457356.14169&-1.975064&1.440952&0.7642  \\
2457379.16194&-4.437901&2.241596&0.6464  \\
2457380.12667&-5.164231&1.315844&0.7863  \\
2457402.09882&1.894179&1.464712&0.7416  \\
2457412.01971&-5.863205&1.562807&0.7138  \\
2457412.97043&-4.887563&1.686561&0.7082  \\
2457440.00883&-0.910402&1.606493&0.8340  \\
2457555.80733&1.681344&1.661285&0.8243  \\
2457556.77194&-2.244875&2.209068&0.7812  \\
2457561.77900&-7.481403&1.4887&0.7161  \\
2457562.79062&-4.631916&1.580454&0.8505  \\
2457568.76731&-10.052094&1.606564&0.7981  \\
2457569.76509&-8.724515&1.558577&0.7550  \\
2457582.75548&-3.797623&1.436803&0.6825  \\
2457583.75403&-3.464112&1.364596&0.6969  \\
2457584.75297&-1.794929&1.603735&0.7044  \\
2457585.75810&-2.151584&1.60559&0.7309  \\
2457586.75440&-3.55797&1.564322&0.3217  \\
2457587.75599&-1.494904&1.358847&0.6771  \\
2457595.74801&10.033845&2.654412&0.3791  \\
2457598.74824&0.027648&2.425191&0.7842  \\
2457704.13227&-0.506104&1.454338&0.6111  \\
2457712.13329&-2.620005&1.708553&0.7164  \\
2457713.13868&-2.609817&1.463878&0.7393  \\
2457748.04174&6.208275&1.448043&0.6792  \\
2457760.06409&-3.215456&1.623777&0.7917  \\
2457764.05803&3.894852&2.171037&0.5624  \\
2457764.07086&-7.417492&1.509814&0.7471  \\
2457766.11836&1.378119&1.697936&0.6227  \\
2457775.02567&-6.355654&1.394235&0.6265  \\
2457776.02209&-6.467462&1.565302&0.6076  \\
2457789.02271&-1.747466&1.475222&0.7479  \\
2457790.06443&-1.473748&1.444261&0.7361  \\
2457790.97319&-5.278218&1.465153&0.6644  \\
2457793.02703&3.394093&1.45068&0.7481  \\
2457794.05774&0.960206&1.40496&0.7684  \\
2457807.14530&-5.723126&1.714786&0.6560  \\
2457830.03452&0.125619&1.574956&0.7394  \\
2457854.95873&-5.315930&1.674391&0.6144  \\
\enddata
\end{deluxetable}

\startlongtable
\begin{deluxetable}{ccccc}
\tablecaption{GJ3470 Relative RV Measurements \label{tab:gj3470rv}}
\setlength\extrarowheight{-1.5pt}
\tablehead{\colhead{BJD$_{TDB}$} & \colhead{RV (m/s)} & \colhead{Unc. (m/s)} & \colhead{$S_\mathrm{HK}$} & \colhead{Instrument}} 
\startdata
2456196.12523 & 7.288214 & 1.593834 & .846 & HIRES\\
2456203.092443 & 8.423508 & 1.834917 & .905 & HIRES\\
2456290.104424 & 6.622361 & 1.963188 & .94 & HIRES\\
2456325.977586 & 2.104225 & 1.866118 & 1.01 & HIRES\\
2456326.975188 & .832374 & 1.758792 & .942 & HIRES\\
2456327.88963 & -7.895415 & 1.692606 & .826 & HIRES\\
2456343.870814 & .923284 & 2.314857 & 1.33 & HIRES\\
2456588.081082 & -.35254 & 1.870567 & 1.13 & HIRES\\
2456589.077408 & .063746 & 1.937277 & 1.15 & HIRES\\
2456614.065159 & 4.259696 & 1.6414 & .94 & HIRES\\
2456638.025333 & .966082 & 1.984093 & 1.003 & HIRES\\
2456639.072967 & 3.538801 & 2.074385 & 1.05 & HIRES\\
2456674.864834 & .191914 & 2.54771 & .5569 & HIRES\\
2456913.131531 & 9.193756 & 1.955105 & 1.011 & HIRES\\
2457057.815422 & 10.436685 & 2.241942 & .5531 & HIRES\\
2457058.750372 & -5.2962 & 2.076247 & .7862 & HIRES\\
2457058.873714 & -8.422584 & 1.941071 & .85 & HIRES\\
2457060.982769 & .678555 & 2.075819 & .9992 & HIRES\\
2457061.033057 & 6.206761 & 2.073267 & 1.228 & HIRES\\
2457061.812464 & -7.524646 & 1.785195 & 1.068 & HIRES\\
2457061.94901 & -6.276048 & 1.787998 & .8942 & HIRES\\
2457291.139283 & 5.543131 & 1.439182 & .7674 & HIRES\\
2457294.128406 & 7.733903 & 1.867845 & 1.421 & HIRES\\
2457295.102025 & 3.158103 & 1.820473 & .9017 & HIRES\\
2457297.077797 & 11.011115 & 1.988343 & 1.009 & HIRES\\
2457327.13922 & 1.997108 & 1.644166 & .2899 & HIRES\\
2457353.078119 & -7.943347 & 2.093493 & .9405 & HIRES\\
2457353.944139 & 4.361439 & 1.818723 & 1.234 & HIRES\\
2457355.060478 & -7.495449 & 1.706797 & 1.067 & HIRES\\
2457356.064562 & -8.723251 & 2.051534 & 1.004 & HIRES\\
2457379.916768 & -4.3828 & 1.77893 & 1.109 & HIRES\\
2457413.787232 & 9.647246 & 2.030947 & .8166 & HIRES\\
2457414.914383 & 7.061204 & 1.988887 & 1.079 & HIRES\\
2457422.800915 & -15.693013 & 1.906881 & .9322 & HIRES\\
2457653.107039 & -4.250508 & 1.991533 & .2492 & HIRES\\
2457654.104405 & 3.586683 & 1.673021 & .9448 & HIRES\\
2457669.122191 & -14.811282 & 1.578996 & 1.148 & HIRES\\
2457672.13511 & -11.112049 & 1.828574 & 1.234 & HIRES\\
2457673.081487 & -13.188282 & 1.909879 & .9718 & HIRES\\
2457679.036588 & -5.932064 & 2.144353 & 1.578 & HIRES\\
2457698.092274 & -1.030799 & 1.614768 & .961 & HIRES\\
2457704.037282 & 2.59695 & 1.831448 & .9315 & HIRES\\
2457715.110143 & -2.959431 & 1.638235 & .8807 & HIRES\\
2457716.043219 & -9.729558 & 1.768124 & 1.031 & HIRES\\
2457717.029436 & -2.234329 & 1.714418 & .9169 & HIRES\\
2457746.949718 & -.646912 & 1.913229 & 1.259 & HIRES\\
2457747.976742 & 11.546778 & 1.926047 & 1.006 & HIRES\\
2457760.011365 & -9.438351 & 1.810585 & 1.073 & HIRES\\
2457761.933534 & -3.068847 & 2.321973 & .8624 & HIRES\\
2457763.825522 & -3.939269 & 1.78412 & 2.341 & HIRES\\
2457774.850285 & 13.127662 & 1.996166 & 1.025 & HIRES\\
2457775.809841 & 2.277267 & 1.93025 & 1.117 & HIRES\\
2457787.841488 & 6.761939 & 1.818927 & .9395 & HIRES\\
2457789.764085 & -11.138971 & 1.723854 & 1.033 & HIRES\\
2457790.753742 & 6.147943 & 1.950918 & 1.051 & HIRES\\
2457828.957103 & -.62745 & 2.03147 & 1.235 & HIRES\\
2455987.609282&26499.6&5.53&--&HARPS\\
2455988.600211&26509.13&3.66&--&HARPS\\
2455989.61836&26520.36&4.12&--&HARPS\\
2455998.576091&26491.47&4.28&--&HARPS\\
2455999.591694&26512.37&4.39&--&HARPS\\
2456000.582575&26493.87&3.92&--&HARPS\\
2456004.556958&26499.35&4.41&--&HARPS\\
2456006.601714&26505.11&3.59&--&HARPS\\
2456009.541376&26508.57&3.51&--&HARPS\\
2456010.556922&26496.39&3.12&--&HARPS\\
2456020.540282&26500.09&3.76&--&HARPS\\
2456021.516266&26484.51&4.64&--&HARPS\\
2456022.539385&26508.62&4.53&--&HARPS\\
2456024.514473&26486.41&3.93&--&HARPS\\
2456026.493726&26505.56&4.99&--&HARPS\\
2456030.52744&26507.06&6.71&--&HARPS\\
2456052.469466&26508.64&3.61&--&HARPS\\
2456053.450485&26502.98&3.56&--&HARPS\\
2456054.474948&26488.64&7.07&--&HARPS\\
2456056.456245&26502.6&3.37&--&HARPS\\
2456058.446999&26494.33&3.22&--&HARPS\\
2456060.445313&26495.35&3.04&--&HARPS\\
2456062.441639&26502.2&3.89&--&HARPS\\
2456253.864979&26500.27&3.54&--&HARPS\\
2456254.863141&26490.16&3.79&--&HARPS\\
2456256.842426&26503.37&4.44&--&HARPS\\
2456258.844023&26497.3&4.5&--&HARPS\\
2456259.821681&26512.3&6.12&--&HARPS\\
2456285.804381&26504.38&4.06&--&HARPS\\
2456286.769862&26505.39&3.95&--&HARPS\\
2456287.774812&26494.21&5.19&--&HARPS\\
2456321.69114&26492.16&4.62&--&HARPS\\
2456360.575823&26506.15&4.55&--&HARPS\\
2456367.577169&26496.9&3.81&--&HARPS\\
2456374.535139&26486.66&3.63&--&HARPS\\
2457689.85429&26493.05&3.56&--&HARPS\\
2457698.86444&26499.26&2.99&--&HARPS\\
2457701.864633&26500.25&3.33&--&HARPS\\
2457703.861325&26501.58&3.4&--&HARPS\\
2457705.84997&26491.55&5.61&--&HARPS\\
2457733.795979&26502.41&4.13&--&HARPS\\
2457787.682997&26508.06&4.1&--&HARPS\\
2457798.632601&26505.13&4.61&--&HARPS\\
2457831.588482&26509.09&3.15&--&HARPS\\
2457832.551386&26496.92&3.76&--&HARPS\\
2457834.591927&26510.64&3.66&--&HARPS\\
2457835.55278&26499.5&5.46&--&HARPS\\
2457836.583106&26491.16&3.86&--&HARPS\\
2457850.508405&26491.16&4.2&--&HARPS\\
2457851.513158&26508.38&4.59&--&HARPS\\
2457852.506325&26494.51&4.07&--&HARPS\\
2457855.51815&26494.85&3.38&--&HARPS\\
\enddata
\end{deluxetable}

An additional 360 Doppler measurements were used in the following K2-3 analysis. We include 31 spectra collected with PFS \citep{Dai2016}, 132 spectra collected with HARPS, and 197 spectra collected with HARPS-N \citep{Almenara2015,Damasso2018}. Our HIRES measurements have an average uncertainty of 1.7 \ms, whereas the PFS, HARPS, and HARPS-N measurements have average uncertainties of 2.5 \ms, 2.1 \ms, and 2.0 \ms\ respectively. An additional 114 Doppler measurements collected with HARPS were used in the following GJ3470 analysis, 61 from the original discovery paper \citep{Bonfils2012} and 53 additional measurements taken in the same fashion \citep{Astudillo2015, Astudillo2017}. Our HIRES measurements have an average uncertainty of 1.9 \ms\ while the two sets of HARPS measurements have an average uncertainty of 4.2 \ms.

\subsection{Stellar Activity}
\label{sec:stellaractivity}

Magnetic activity on the stellar surface can induce planet-like signals in RV data \citep{Robertson2013,Robertson2015}. This is especially problematic for M dwarfs, where the magnetic activity is not as well characterized as for solar-type stars and the stellar rotation period is often similar to planet orbital periods at days to tens of days \citep{McQuillan2013,Newton2016}. The stellar activity also causes absorption line variability \citep{Cincunegui2007,Buccino2011,Silva2012}, which can be tracked by measuring tracers such as the Calcium II H and K lines, noted as $S_\mathrm{HK}$. $S_\mathrm{HK}$ may not always indicate activity for M dwarfs \citep{Robertson2015}; both photometry and H$\alpha$ can be useful diagnostics for M-dwarf stellar rotation periods \citep{Newton2017} . 

We first examined the potential effects of stellar activity by measuring the strength of these Calcium II H and K spectral lines in our HIRES RV measurements \citep{Isaacson2010}. We calculated the correlation coefficient and probability value ($p$-value) for the $S_\mathrm{HK}$ and RV data for each season of data collection (using scipy, \citet{Scipy}). 
Then, we examined the RV and $S_\mathrm{HK}$ periodograms for potential similarities. We also analyzed ground-based photometry of K2-3 and GJ3470 to determine the rotation period and compared this period to the RV periodograms. 
Finally, we modeled the RV data of K2-3 and GJ3470 with Gaussian processes (GPs) trained on the photometry to remove correlated noise in the RVs from the stellar activity. 

\subsubsection{K2-3 Stellar Activity and Ground-based Photometry}
\label{sec:k2phot}

\begin{figure}[!h]
\hspace*{-0.4cm} 
\includegraphics[trim={0 6cm 0 0},clip,width=0.5\textwidth]{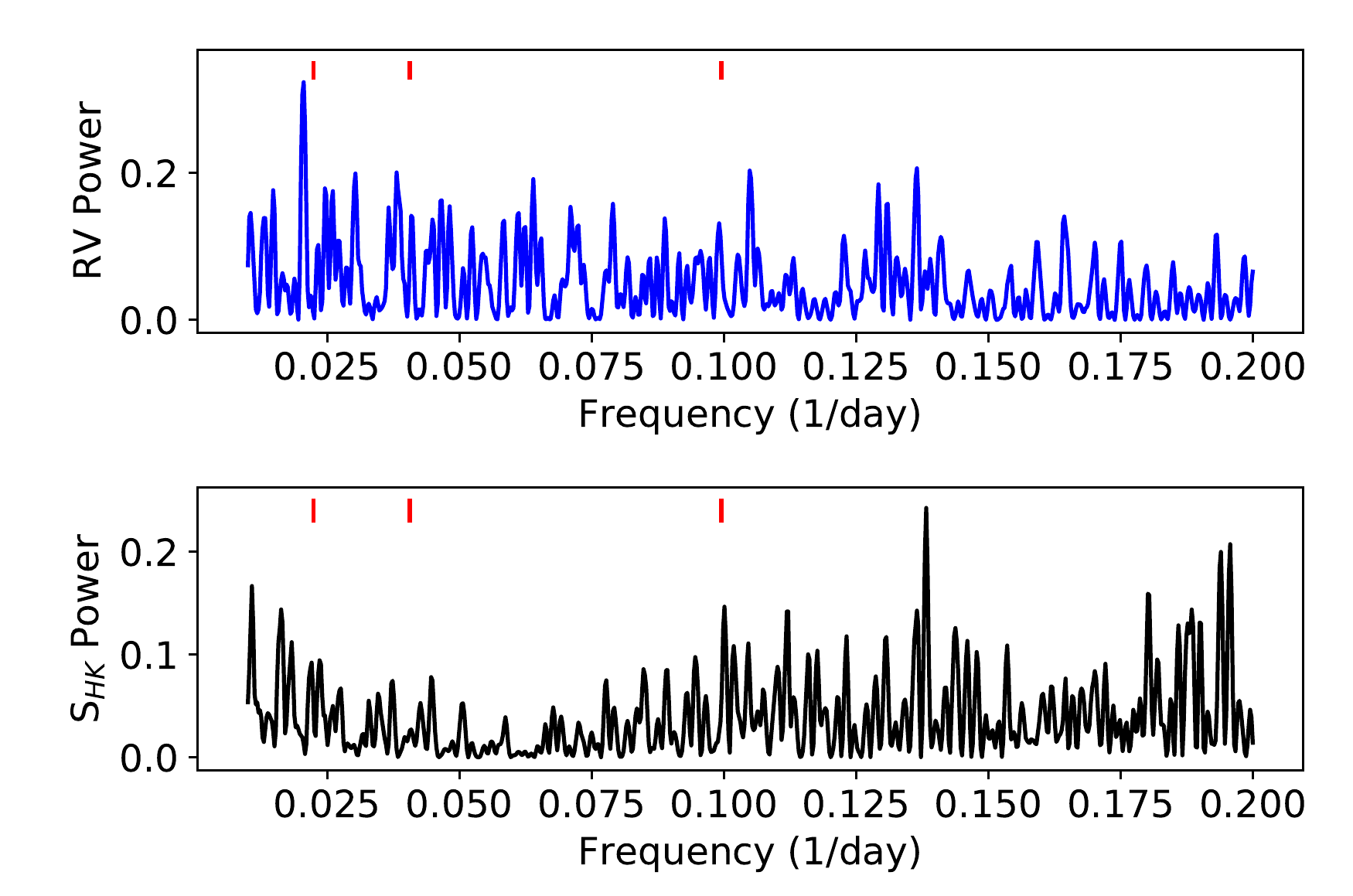}
\caption{K2-3: periodograms of the radial velocity data (top) and $S_\mathrm{HK}$ (bottom). The three planet periods are shown by red tick marks at the top of the figures. The RV and $S_\mathrm{HK}$ periodogram do not have similar prominent peaks. Although the planet periods are not all visible in the RV periodogram due to their meter per second RV amplitudes, we are able to detangle the planet signals in the RV data by constraining the periods and conjunction times from \kt\ and \spitzer\ transits. 
\label{fig:shkperiod}
}
\end{figure}

We investigate the possible correlation between $S_\mathrm{HK}$ and RV values for K2-3 (\autoref{tab:k23rv}) as the stellar rotation period found from \kt\ photometry \citep[40 $\pm$ 10 days;][]{Dai2016} is near the orbital period of planet d. \citet{Dai2016} and \citet{Damasso2018} find the planet signal to be degenerate with the stellar rotation signal. 
The correlation coefficient is -0.0169 and $p$-value is 0.8869 for the full dataset, suggesting that the RVs are not correlated with the stellar activity as measured by $S_\mathrm{HK}$. 
We also do not find any similar significant peaks in the periodograms (\autoref{fig:shkperiod}). However, as the RV periodogram does not show all of the planet signals, the activity signals may also be hidden.

\begin{figure}[ht]
%\centering
\hspace*{-1cm}  
\includegraphics[width=3.8in]{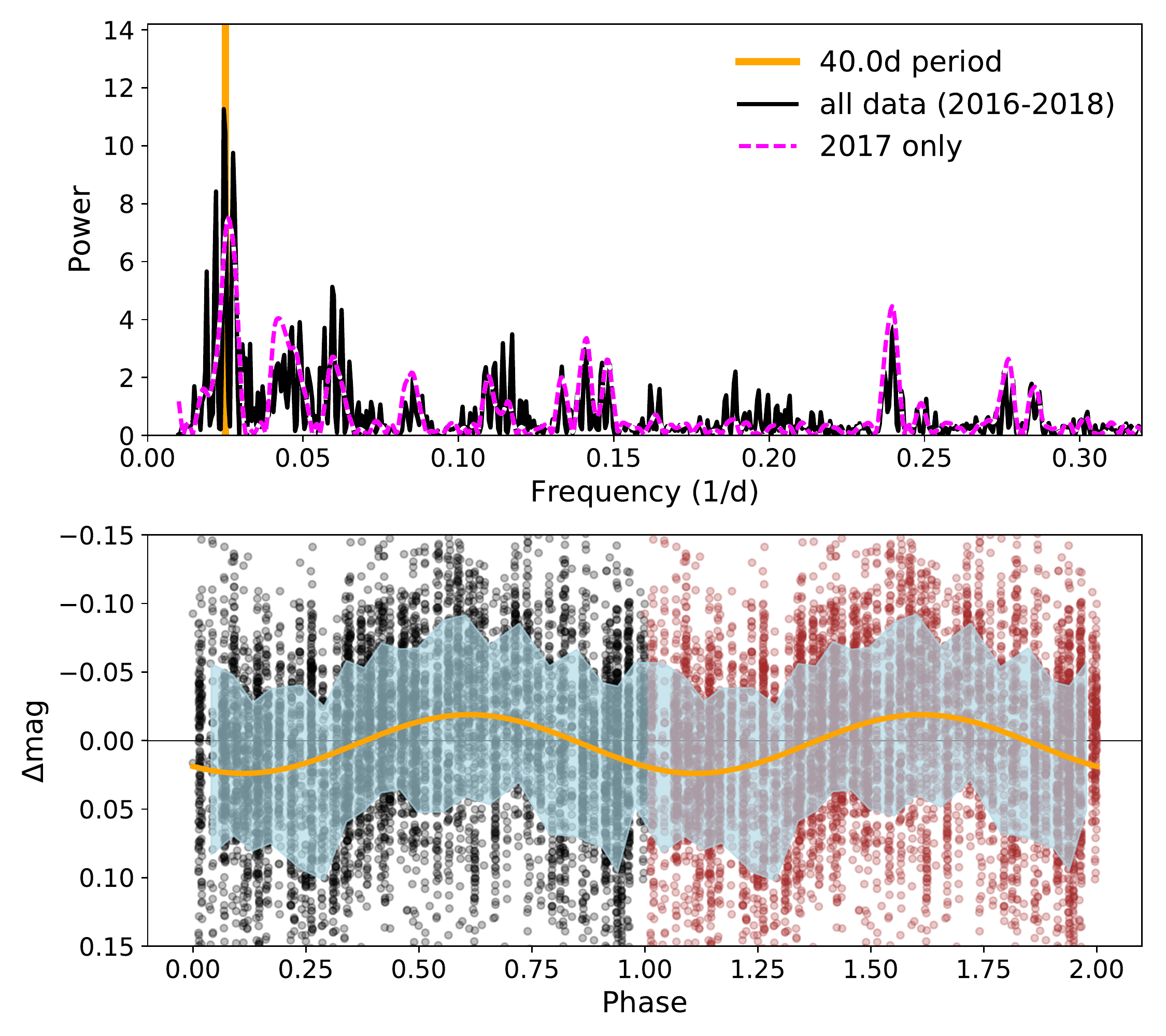}
\caption{
Evryscope photometry of K2-3, consisting of 9931 epochs at 2 minute cadence in Sloan-g' from 2016 Jan to 2018 March. Top panel: the LS periodogram of K2-3 displays significant power around 40 days (orange line). A purple dashed line shows the power of only the 2017 photometry as a secondary confirmation. Bottom panel: phase-folded light curve folded over 40 days. The phase is repeated to guide the eye, and points are binned to eight-minute cadence to improve precision on this relatively faint Evryscope target. The 1$\sigma$ region about the mean of the phased lightcurve is shown (light blue area), along with a 40-day sinusoid with a characteristic amplitude of 0.02 mag (orange curve). 
\label{fig:evryscope}
}
\end{figure}

As mentioned above, $S_\mathrm{HK}$ may be a poor indicator for M-dwarf stars. To better characterize the possible rotation signal of K2-3, we analyzed photometry from the Evryscope. The Evryscope is an array of 24 61mm telescopes together imaging 8000 square degrees of sky every two minutes \citep{Law2015}. Since its 2015 installation at Cerro Tololo Inter-American Observatory (CTIO) in Chile, the Evryscope has observed on over 500 clear nights, tracking the sky for two hours at a time before ratcheting back and continuing observations, for an average of $\sim$6 hours of continuous monitoring each night. The Evryscope observes in Sloan-$g$' at a resolution of 13"/pixel. High-cadence photometry of K2-3 is included in the Evryscope light curve database from 2016 January to 2018 March (\autoref{fig:evryscope}). Because K2-3 is in the northernmost region of the Evryscope field of view, the coverage of the target is limited each year, resulting in a total of $10^4$ epochs; most southern stars are observed with 4--6$\times$ more points.

Evryscope light curves are generated using a custom pipeline. The Evryscope image archive contains 2.5 million raw images, $\sim$250TB of total data. Each image, consisting of a 30MPix FITS file from one camera, is dark-subtracted, flat-fielded and then astrometrically calibrated using a custom wide-field solver. Large-scale background gradients are removed, and forced-aperture photometry is then extracted based on known source positions in a reference catalog. Light curves are generated for approximately 15 million sources across the southern sky by differential photometry in small sky regions using carefully selected reference stars; residual systematics are removed using two iterations of the SysRem detrending algorithm. For 10th mag stars, this process results in $\approx$1\% photometric stability at two-minute cadence when measured in multiple-year light curves over all sky conditions; co-adding produces improved precisions, down to $\sim$6 mmag.

Evryscope collected 9931 epochs of K2-3 at two-minute cadence in Sloan-$g$' from 2016 January 2016 to 2018 March. The data were analyzed using a Lomb-Scargle (LS) periodogram to determine the likely rotation period of K2-3 (\autoref{fig:evryscope}). The highest peak is at 40.0 days, but power from the central peak is split due to the inter-year window function, verified by injecting similar signals to K2-3 and other nearby stars. An alias of the 40.0 day signal exists at a reduced power near 20 days. The periodogram for only the 2017 photometry produces a peak signal of 38 days. A signature of evolving starspot activity due to differential rotation near 40 days may explain this difference. The 40.0 day period shows a sinusoidal variation with a 0.02 mag variation.  Therefore, we infer the rotation period of K2-3 to be \krotper\ days from the Evryscope data. Both the 2017 and the all-data rotation periods agree with the estimate from \kt\ data, within measurement errors. We use the Evryscope photometry to inform our GP priors in the RV fit (Section \ref{sec:k2rv}). 

\subsubsection{GJ3470 Stellar Activity}
\label{sec:gjrot}

\begin{figure}[ht]
\hspace*{-0.5cm}
\includegraphics[width=3.7in]{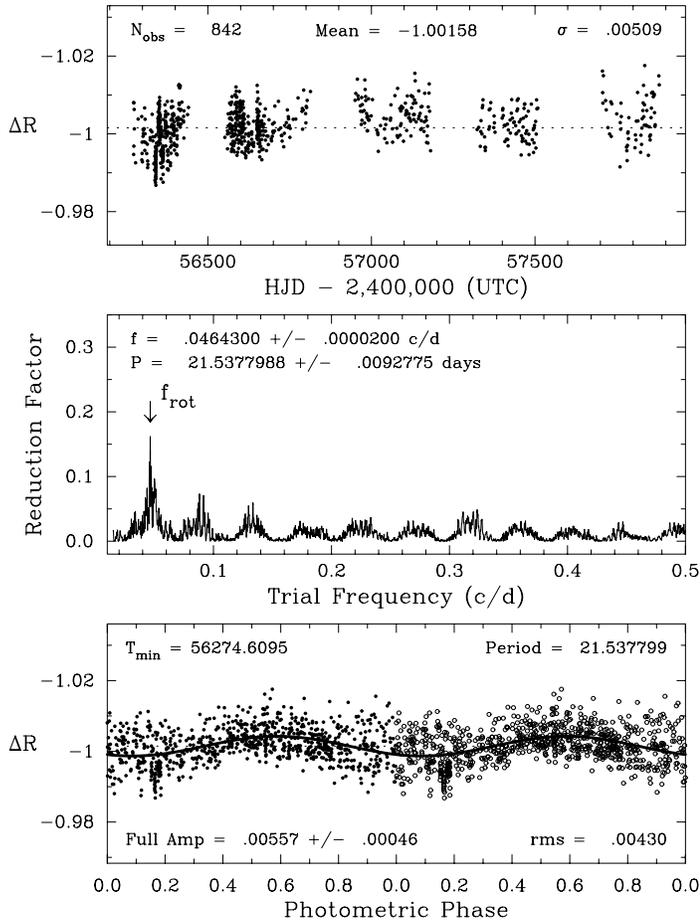}
\caption{Top: Photometry of GJ3470 from 2012 to 2017 from the C14 AIT at Fairborn Observatory. Middle: Power spectrum of the observations in frequency space resulted in a stellar rotation period of 21.54 days. We inflated the period uncertainty to \rotper\ days, to account for the variation in rotation period over time. Bottom: Phased photometry over the periodogram peak at 21.54 days.}
\label{fig:gjphotperiod}
\end{figure}

\begin{deluxetable}{ccccc}
\centering
\tablecaption{Summary of C14 AIT Photometric Observations of GJ3470 \label{tab:FO}}
\tablehead{\colhead{Observing} & \colhead{Date Range} & \colhead{} & 
\colhead{Sigma} & \colhead{Seasonal Mean} \\
\colhead{Season} & \colhead{(HJD$-$2,400,000)} & \colhead{$N_{obs}$} &
\colhead{(mag)} & \colhead{(mag)} }
\startdata
2012-2013 & 56272--56440 & 297 & 0.00535 & $-0.99917\pm0.00031$ \\
2013-2014 & 56551--56813 & 289 & 0.00397 & $-1.00205\pm0.00023$ \\
2014-2015 & 56949--57180 & 108 & 0.00419 & $-1.00494\pm0.00040$ \\
2015-2016 & 57323--57508 &  83 & 0.00384 & $-1.00214\pm0.00042$ \\
2016-2017 & 57705--57879 &  65 & 0.00586 & $-1.00417\pm0.00073$ \\
\enddata
\end{deluxetable}

\begin{figure}[ht]
\centering
\vspace{0.5cm}
\hspace*{-0.4cm} 
\includegraphics[trim={0 6cm 0 0},clip,width=0.5\textwidth]{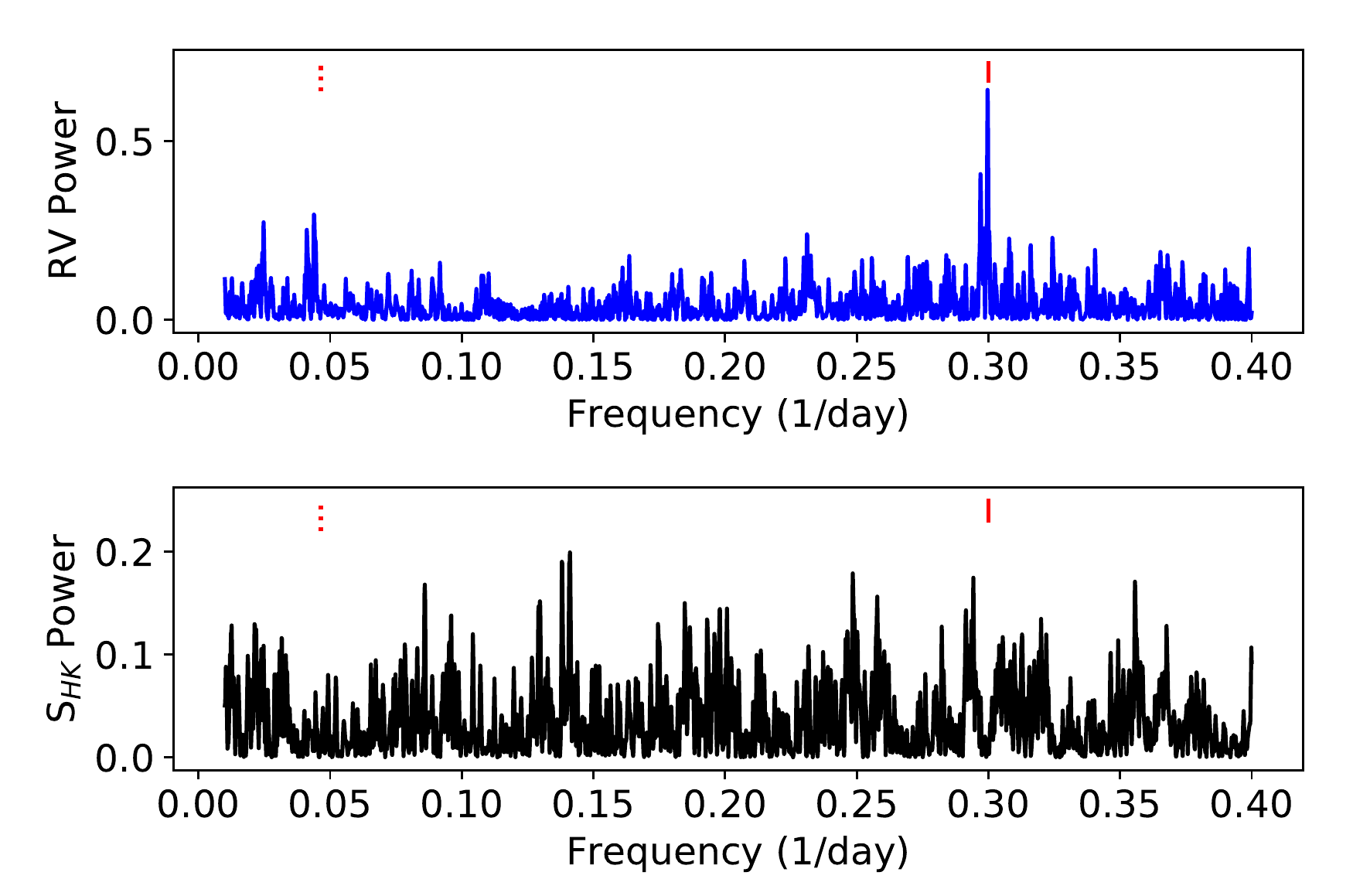}
\caption{GJ3470: periodogram of the radial velocity data (top) and $S_\mathrm{HK}$ periodogram (bottom). The main peak in RV at 3.3 days matches the period of planet b (red tick mark). 
%The second peak at 1.42 days is an alias of the planet period and the 1-day time sampling \citep{Bonfils2012}. 
The next prominent peaks are near the stellar rotation period (red dotted mark; see~\autoref{fig:gjphotperiod}). The RV and $S_\mathrm{HK}$ periodograms do not have any prominent peaks in common.} 
\label{fig:gjrvvsshk}
\end{figure}

To better characterize GJ3470, photometry was collected at the Fairborn Observatory in Arizona with the Tennessee State University Celestron C14 0.36 m Automated Imaging Telescope (AIT) \citep{Henry1999,Eaton2003}. The AIT has a SBIG STL-1001E CCD camera and a Cousins R filter. Images were corrected for bias, flat-fielding, and differential extinction. Differential magnitudes were computed using five field stars. 842 observations were collected from 2012 December to 2017 May (\autoref{tab:FO}, \autoref{fig:gjphotperiod}).

The tallest peak in the periodogram (\autoref{fig:gjphotperiod}) corresponds to a period of \rotper\ days; the uncertainty is the standard deviation between the peaks for each observing season. We interpret this peak as the stellar rotation period, as shown by the brightness variation from star spots rotating in and out of view. This rotation period is consistent with that found by \citet{Biddle2014}. 

For GJ3470, there is a hint of an RV-$S_\mathrm{HK}$ correlation in the early HIRES data, although the full dataset has a correlation coefficient of -0.0753 and $p$-value of 0.5812. Furthermore, the RV periodogram contains a significant peak near the stellar rotation period (\autoref{fig:gjrvvsshk}), which suggests that the stellar rotation signal needs to be accounted for in the RV analysis. We therefore used this photometry to inform our GP priors in the RV fit (Section~\ref{sec:gjrv}).

\subsection{RV Analysis}
\label{sec:rv}

We analyzed the RV data for both systems using RadVel, an open-source orbit-fitting toolkit for RV data \citep{Fulton2018}. RadVel models the RVs as the sum of Keplerian orbits. The model parameters are orbital period (P), time of inferior conjunction (T$_{conj}$), radial velocity amplitude (K), eccentricity ($e$), argument of periastron ($\omega$), a constant RV offset ($\gamma$), and a jitter term for each instrument ($\sigma$). In order to avoid biasing eccentricity, $\sqrt{e}cos(\omega)$ and $\sqrt{e}sin(\omega)$ are used as fitting parameters.

We modeled the correlated noise introduced from the stellar activity using a quasi-periodic Gaussian Process (GP) with a covariance kernel of the form 
\begin{equation}
\label{eq:kernel}
    k(t,t') = \eta_1^2 \ \rm{exp} \left[-\frac{(t-t')^2}{\eta_2^2}-\frac{sin^2(\frac{\pi(t-t')}{\eta_3})}{\eta_4^2})\right],
\end{equation}
where the hyper-parameter $\eta_1$ is the amplitude of the covariance function, $\eta_2$ is the active region evolutionary time scale, $\eta_3$ is the period of the correlated signal, and $\eta_4$ is the length scale of the periodic component \citep{LopezMorales2016,Haywood2014}. We trained these parameters on the ground-based photometry of each star by performing a maximum likelihood fit to the associated ground-based light curve with the quasi-periodic kernel (Eq.~\ref{eq:kernel}) then determined the errors through a MCMC analysis. We then compare the period of the correlated signal ($\eta_{3}$) with the stellar rotation period found from our periodogram analysis in Section~\ref{sec:stellaractivity}.

\subsubsection{GJ3470 RV Analysis}
\label{sec:gjrv}

For our RV analysis of GJ3470, we adopt the period, time of conjunction, and planet radius derived from a variety of ground-based telescopes \citep{Biddle2014}. The remaining parameters were initialized from \citet{Dragomir2015}. 

We used a GP to model the correlated noise associated with the stellar activity in our RV fit. We ran our GP analysis on the photometry from Fairborn Observatory (FO, Section~\ref{sec:stellaractivity}) and find $\gamma_{FO}$ = $1.003\pm0.001$, $\sigma_{FO}$ = $0.0029\pm0.0001$, $\eta_1$ = $-0.0036^{+0.0003}_{-0.0004}$, $\eta_2$ = $48.98^{+9.54}_{-7.28}$, $\eta_3$ = $21.84^{+0.35}_{-0.36}$, and $\eta_4$ = $0.55\pm0.06$. This stellar rotation period ($\eta_3$) is consistent with the results of our periodogram analysis in Section~\ref{sec:stellaractivity} to within 1$\sigma$. 

We then perform our RV fit including a GP modeled as a sum of two quasi-periodic kernels, one for each instrument, as HIRES and HARPS have different properties that could alter the effect of stellar activity on the data. Each kernel includes identical $\eta_2$, $\eta_3$, and $\eta_4$ parameters but allows for different $\eta_1$ values. Our priors are as follows: $\eta_1$ is left as a free parameter as light curve amplitude cannot be directly translated to RV amplitude, for $\eta_2$, $\eta_3$, and $\eta_4$, we used a kernel density estimate (KDE) of the Fairborn Observatory photometry posteriors.

After running an initial RV fit including only one circular, Keplerian planet signal, we investigated models including an acceleration term, curvature term, and eccentricity. The Aikike information criterion (AIC) was used to determine if the fit improvement justified the additional parameters; a $\Delta$AIC of $<$ 2 indicates a similar fit, 2 $<$ $\Delta$AIC$<$ 10 favors the additional parameter, and a $\Delta$AIC $>$ 10 is a strong justification for the additional parameter. Only the eccentricity parameters improved the AIC ($\Delta$AIC$_{acc}$ = -0.71, $\Delta$AIC$_{curv}$ = -1.44, and $\Delta$AIC$_{ecc}$ = 6.45).
All of the tested RV models resulted in planet masses within 1$\sigma$ of the circular fit values shown in~\autoref{tab:gjparams} %\st{and Figure 7}.

\begin{figure*}[!h]
\centering
\includegraphics[height=8.0in,width=6.0in,keepaspectratio]{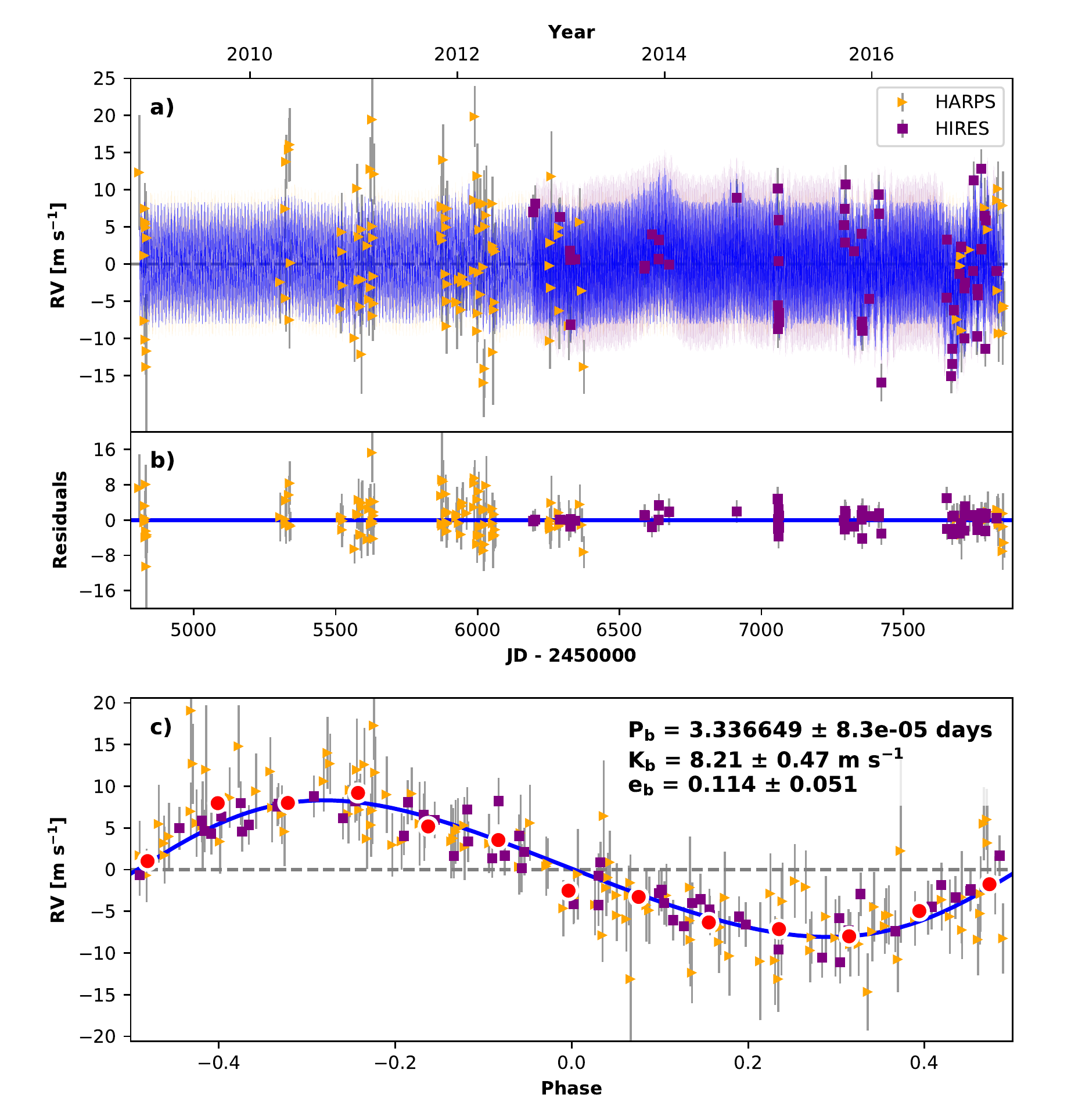}
\vspace{-10pt}
\caption{
Best-fit one-planet Keplerian orbital model for GJ3470 with $e$cos$(\omega)$ constraints from the secondary eclipse observation. The maximum likelihood model is plotted while the orbital parameters listed in~\autoref{tab:gjparams} are the median values of the posterior distributions. The thin blue line is the best fit one-planet model with the mean GP model; the colored area surrounding this line includes the 1$\sigma$ maximum likelihood GP uncertainties. We add in quadrature the RV jitter term(s) listed in~\autoref{tab:gjparams} with the measurement uncertainties for all RVs. {\bf b)} Residuals to the best-fit one-planet model. {\bf c)} RVs phase-folded to the ephemeris of planet b.  The small point colors and symbols are the same as in panel {\bf a)}. The red circles are the same velocities binned in 0.08 units of orbital phase.
\label{fig:gjrv}
}
\end{figure*}

We then investigated a model including an eccentricity constraint from \spitzer\ observations of the secondary eclipse. The secondary eclipse was 0.309 days later than expected for a circular orbit, which results in a constraint on $e$cos$(\omega)$ of $0.014546^{+0.000753}_{-0.000659}$ \citep{Bennekeprep}. For this fit we used $e$cos$(\omega)$ and $e$sin$(\omega)$ as the fitting basis due to the prior set by the secondary eclipse. We find an eccentricity of $e_b = 0.114 \pm 0.051$ for the eccentric model constrained by this secondary eclipse measurement, the best-fit curve is shown in~\autoref{fig:gjrv}.

The non-zero eccentricity value of GJ3470 b is particularly interesting in the context of other systems. GJ436 b, another planet similar in mass, radius, period, and stellar host, has a puzzlingly high eccentricity of 0.150 $\pm$ 0.012 \citep{Deming2007}. These high eccentricity values may be an emerging clue on how these types of planets form and migrate. 

\begin{deluxetable}{lcl}
\tablecaption{GJ3470 RV MCMC Priors and Posteriors \label{tab:gjparams}}
\setlength\extrarowheight{-1pt}
\tablehead{\colhead{Parameter} & \colhead{Value} & \colhead{Units} }
\startdata
\sidehead{\bf{Gaussian Priors}}
$T\rm{conj}_{b}$ & $2455953.6645 \pm 0.0034$ & JD \\
$P_{b}$ &  $3.3371 \pm 0.0002$& days\\
$e\cos{\omega}_{b}$ &  $0.01454 \pm 0.00075323$ & \\
$\eta_{1,HIRES}$ & [0,100] & \ms \\
$\eta_{1,HARPS}$ & [0,100] & \ms \\
$\eta_{2}$ & $48.98^{+9.54}_{-7.28}$ & days \\
$\eta_{3}$ &  $21.84^{+0.35}_{-0.36}$ & days \\
$\eta_{4}$ & $0.55 \pm 0.006$ & \\
\hline
\sidehead{\bf{Orbital Parameters}}
$P_{b}$ & $3.336649^{+8.4e-05}_{-8.1e-05}$ & days\\
$T\rm{conj}_{b}$ &  $2455953.663\pm 0.0035$ & JD\\
$e_{b}$ & $0.114^{+0.052}_{-0.051}$ & \\
$\omega_{b}$ & $-1.44^{+0.1}_{-0.04}$ & radians\\
$K_{b}$ & $8.21^{+0.47}_{-0.46}$ & m s$^{-1}$\\
$M_{b}$ & $12.58^{+1.31}_{-1.28}$  &  M$_{\oplus}$ \\
$\rho_{b}$ & $0.93^{+0.56}_{-0.31}$  & g cm$^{-3}$ \\
\hline
\sidehead{\bf{Other Parameters}}
$\gamma_{\rm HIRES}$ & $0.3^{+1.2}_{-1.1}$ &m s$^{-1}$\\
$\gamma_{\rm HARPS}$ & $26500.52^{+0.59}_{-0.6}$ & m s$^{-1}$\\
$\dot{\gamma}$ & $\equiv$ 0.0  & m s$^{-1}$ day$^{-1}$\\
$\ddot{\gamma}$ & $\equiv$ 0.0  & m s$^{-1}$ day$^{-2}$\\
$\sigma_{\rm HIRES}$ & $1.9^{+0.7}_{-0.67}$ & $\rm m\ s^{-1}$\\
$\sigma_{\rm HARPS}$ &  $0.0023^{+0.49}_{-0.0023}$  & $\rm m\ s^{-1}$\\
$\eta_{1, HIRES}$ & $3.94^{+0.90}_{-0.78}$ & \ms \\
$\eta_{1, HARPS}$ & $1.79^{+0.69}_{-0.71}$  & \ms \\
$\eta_{2}$ & $49.40^{+10.00}_{-7.55}$   & days \\
$\eta_{3}$ & $21.92^{+0.42}_{-0.41}$  &  days \\
$\eta_{4}$ & $0.56 \pm 0.06$  & \\
\enddata
\end{deluxetable}

\subsubsection{K2-3 RV analysis}
\label{sec:k2rv}

For our RV analysis of K2-3, we adopt the planet orbital periods and times of conjunction from our \spitzer\ analysis (Section~\ref{sec:spitzer}). We used a GP to model the correlated noise associated with the stellar activity in our RV fit. We ran our GP analysis on the photometry from Evryscope (ES, Section~\ref{sec:stellaractivity}) and find $\gamma_{ES}$ = $11.61\pm0.01$, $\sigma_{ES}$ = $0.017^{+0.004}_{-0.003}$, $\eta_1$ = $0.03\pm0.01$, $\eta_2$ = $44.57^{+12.58}_{-16.23}$, $\eta_3$ = $37.80^{+1.77}_{-2.04}$, and $\eta_4$ = $0.47\pm0.05$. This stellar rotation period ($\eta_3$) is consistent with the results of our periodogram analysis in Section~\ref{sec:stellaractivity} to within 2-$\sigma$. 

We then perform our RV fit including a GP modeled as a sum of four quasi-periodic kernels, one for HIRES, HARPS, HARPS-N and PFS, as described above for GJ3470.
Our GP hyperparameter priors are as follows: $\eta_1$ is left as a free parameter as light curve amplitude cannot be directly translated to RV amplitude. To construct priors on $\eta_2$, $\eta_3$, and $\eta_4$, we use a KDE of the Evryscope photometry posteriors. 

\begin{figure*}[!h]
\centering
\includegraphics[width=4.5in]{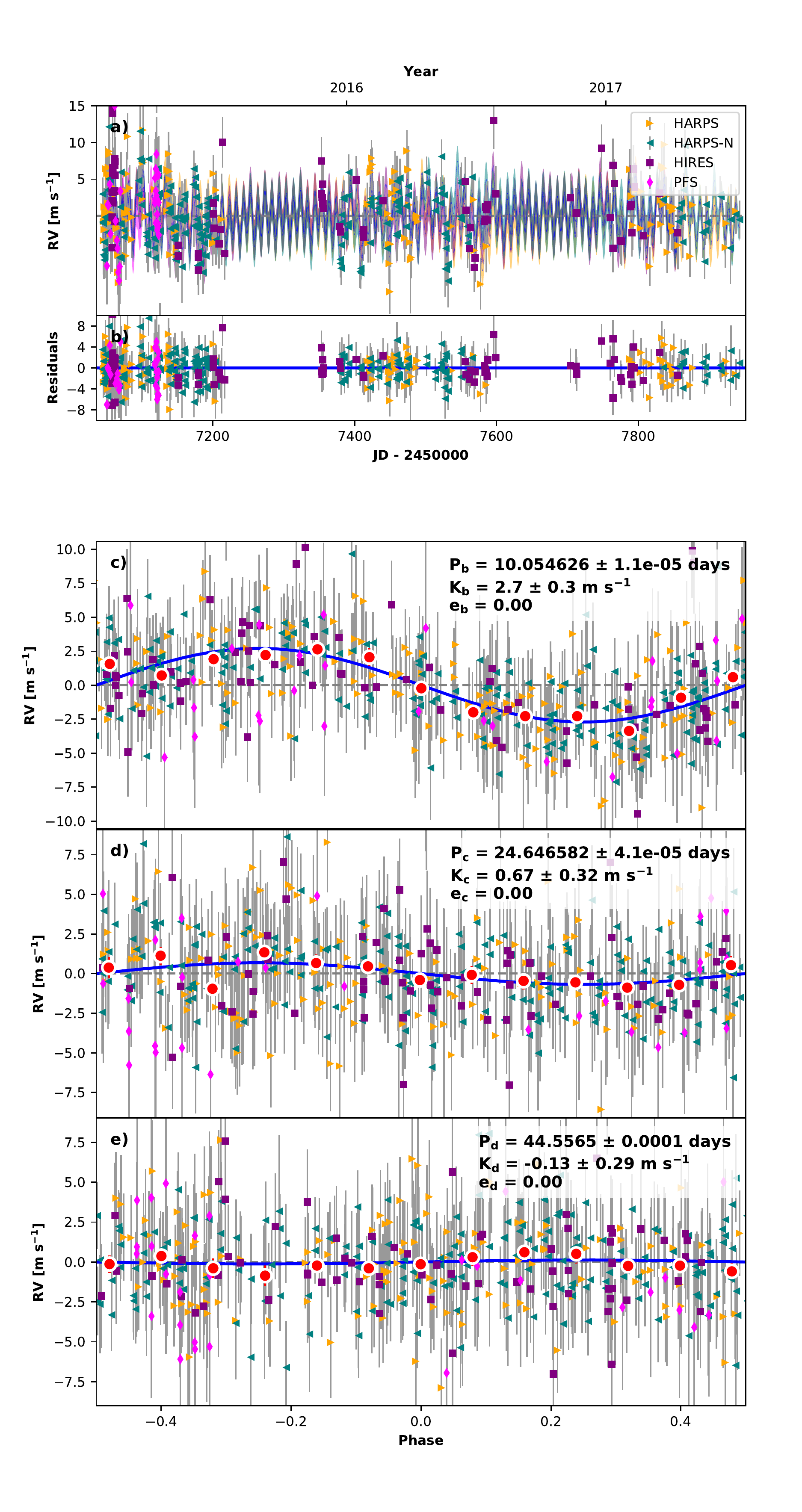}
\vspace{-10pt}
\caption{ 
Best-fit three-planet Keplerian orbital model for K2-3. The maximum likelihood model is plotted while the orbital parameters listed in~\autoref{tab:k23params} are the median values of the posterior distributions. The thin blue line is the best-fit three-planet model with the mean GP model; the colored area surrounding this line includes the 1-$\sigma$ maximum likelihood GP uncertainties. We add in quadrature the RV jitter term(s) listed in~\autoref{tab:k23params} with the measurement uncertainties for all RVs. {\bf b)} Residuals to the best-fit three-planet model. {\bf c)} RVs phase-folded to the ephemeris of planet b. The Keplerian orbital models for all other planets have been subtracted. The small point colors and symbols are the same as in panel {\bf a)}. The red circles are the same velocities binned in 0.08 units of orbital phase. The phase-folded model for planet b is shown as the blue line. Panel {\bf d)} and {\bf e)} are the same as panel {\bf c)} but for planet K2-3 c and d, respectively.
\label{fig:rvcirc}
}
\end{figure*}

After running an initial RV fit including only three circular, Keplerian planet signals, we investigated additional models including an acceleration term, curvature term, and planet eccentricity. The fit including an additional term for acceleration, curvature, and eccentricity had $\Delta$AIC of -2.17, -2.22, and -2.92 respectively; none of these justified the additional parameter. ~\autoref{tab:k23params} shows the MCMC priors, orbital parameters, and statistics for the GP model of K2-3. The best-fit curves for each planet is shown in~\autoref{fig:rvcirc}.

\startlongtable
\begin{deluxetable}{lcr}
\setlength{\tabcolsep}{0pt}
\tablecaption{K2-3 RV MCMC Priors and Posteriors \label{tab:k23params}}
\setlength\extrarowheight{-1pt}
\tablehead{\colhead{Parameter} & \colhead{Three-Planet Fit} & \colhead{Units}}
\startdata 
\sidehead{\bf{Gaussian Priors}}
$T\rm{conj}_{b}$ &$2456813.41843 \pm 0.00039$& JD\\
$P_{b}$ &$10.054626 \pm 1e-05$ & days\\
$T\rm{conj}_{c}$ & $2456812.28013 \pm 0.00095$ & JD\\
$P_{c}$ &  $24.646582 \pm 3.9e-05$& days\\
$T\rm{conj}_{d}$ & $2456826.22347 \pm 0.00053$ & JD\\
$P_{d}$ & $44.556456 \pm 9.7e-05$ & days\\
$\eta_{1,all}$ & [0,100] & \ms \\
$\eta_{2}$ & $44.57^{+12.58}_{-16.23}$ & days\\
$\eta_{3}$ & $37.80^{+1.77}_{-2.04}$ & days\\
$\eta_{4}$ & $0.47\pm 0.05$ & \\
\hline
\sidehead{\bf{Orbital Parameters}}
$P_{b}$ & $10.054626^{+1e-05}_{-1.1e-05}$  & days\\
$T\rm{conj}_{b}$ &$2456813.41843\pm 0.00041$  & JD\\
$e_{b}$ & $\equiv$ 0.0  & \\
$\omega_{b}$ & $\equiv$ 0.0  & radians\\
$K_{b}$ & $2.72^{+0.29}_{-0.3}$& m s$^{-1}$\\
$M_{b}$ & $6.48^{+0.99}_{-0.93}$ & \mearth \\ 
$\rho_{b}$ & $3.70^{+1.67}_{-1.08}$ & g cm$^{-3}$ \\
$P_{c}$ & $24.646582^{+4.1e-05}_{-4e-05}$  & days\\
$T\rm{conj}_{c}$ & $2456812.28018^{+0.00098}_{-0.001}$  & JD\\
$e_{c}$ & $\equiv$ 0.0  &  \\
$\omega_{c}$ & $\equiv$ 0.0  &  radians\\
$K_{c}$ &   $0.67\pm 0.32$  & m s$^{-1}$\\
$M_{c}$ & $2.14^{+1.08}_{-1.04}$ & \mearth \\
$\rho_{c}$ & $2.98^{+1.96}_{-1.50}$ & g cm$^{-3}$ \\
$P_{d}$ &  $44.55646^{+0.00011}_{-0.0001}$   &  days\\
$T\rm{conj}_{d}$ & $2456826.22346\pm 0.00056$ & JD\\
$e_{d}$ & $\equiv$ 0.0  &  \\
$\omega_{d}$ & $\equiv$ 0.0  & radians\\
$K_{d}$ &  $-0.13^{+0.28}_{-0.31}$ & m s$^{-1}$\\
$M_{d}$ & $-0.50^{+1.10}_{-1.20}$ & \mearth \\
$\rho_{d}$ & $-0.98^{+2.20}_{-2.83}$ & g cm$^{-3}$ \\
$K_{d}$  (3$\sigma$ upper) & $0.71$  & m s$^{-1}$\\
$M_{d}$ (3$\sigma$ upper) & $2.80$ & \mearth \\
$\rho_{d}$ (3$\sigma$ upper) & $5.62$ & g cm$^{-3}$ \\
\hline
\sidehead{\bf{Other Parameters}}
$\gamma_{\rm PFS}$ &  $-1.3\pm 2.2$  & m s$^{-1}$\\
$\gamma_{\rm HIRES}$ &  $-2.98^{+0.97}_{-1.0}$ &  m s$^{-1}$\\
$\gamma_{\rm HARPS-N}$ &  $0.53^{+0.71}_{-0.74}$ & $\rm m\ s^{-1}$ \\
$\gamma_{\rm HARPS}$ & $-0.59^{+0.69}_{-0.73}$ &  m s$^{-1}$\\
$\dot{\gamma}$ & $\equiv$ 0.0  &  m s$^{-1}$ day$^{-1}$\\
$\ddot{\gamma}$ & $\equiv$ 0.0  &  m s$^{-1}$ day$^{-2}$\\
$\sigma_{\rm PFS}$ &  $4.85^{+1.0}_{-0.88}$&  $\rm m\ s^{-1}$\\
$\sigma_{\rm HIRES}$ &  $2.98^{+0.47}_{-0.42}$ & $\rm m\ s^{-1}$\\
$\sigma_{\rm HARPS-N}$ &  $1.61^{+0.26}_{-0.25}$  & $\rm m\ s^{-1}$ \\
$\sigma_{\rm HARPS}$ & $2.06^{+0.34}_{-0.32}$ & $\rm m\ s^{-1}$\\
$\eta_{1,PFS}$ & $4.75^{+3.72}_{-2.58}$ & m s$^{-1}$ \\
$\eta_{1,HIRES}$ & $3.21^{+0.84}_{-0.73}$ & m s$^{-1}$ \\
$\eta_{1,HARPS}$ & $3.04^{+0.64}_{-0.53}$ & m s$^{-1}$ \\
$\eta_{1,HARPS-N}$ &$3.07^{+0.61}_{-0.48}$ & m s$^{-1}$ \\
$\eta_{2}$ & $62.25^{+10.78}_{-9.84}$ & days \\
$\eta_{3}$ & $39.16^{+0.88}_{-0.96}$ & days \\
$\eta_{4}$ & $0.41^{+0.05}_{-0.04}$ & \\
\enddata
\end{deluxetable}

From our GP fit, we find that the semi-amplitude of the signal from planet d is consistent with 0 to 1$\sigma$. It is possible that this planet has a small semi-amplitude (K$_d$ $<<$ \ms) and we were unable to detect it. Alternatively, as the period of planet d (P$_d$ = 44.56 days) is near the stellar rotation period ($\eta_3 \approx$ 40 days), it is possible that the signal of planet d is indistinguishable from the stellar activity signal. Further work is needed to distinguish between the two possibilities and determine the mass of planet d.

\section{Discussion}
\label{sec:disc}
These four Earth- to Neptune-sized planets are great candidate targets for atmospheric transmission spectroscopy due to their bright host stars (K $<$ 9 mag) and low densities ($<4.2$ g cm$^{-3}$). GJ3470 b has been observed with the \textit{Hubble Space Telescope} (HST) in cycles 19 and 22 (GO 13064, GO 13665). K2-3 d and the K2-3 UV emission will be observed in cycles 24 and 25 (GO 14682, GO 15110). GJ3470 b already shows H$_2$O absorption (\citet{Tsiaras2017}; Benneke et al. 2018, submitted) and is being targeted by JWST Guaranteed Time Observation (GTO) program observations. 
%https://archive.stsci.edu/proposal_search.php?mission=hst&id=14682 search feature. 

\begin{figure}[h]
\hspace*{-1cm}  
\includegraphics[width=4in]{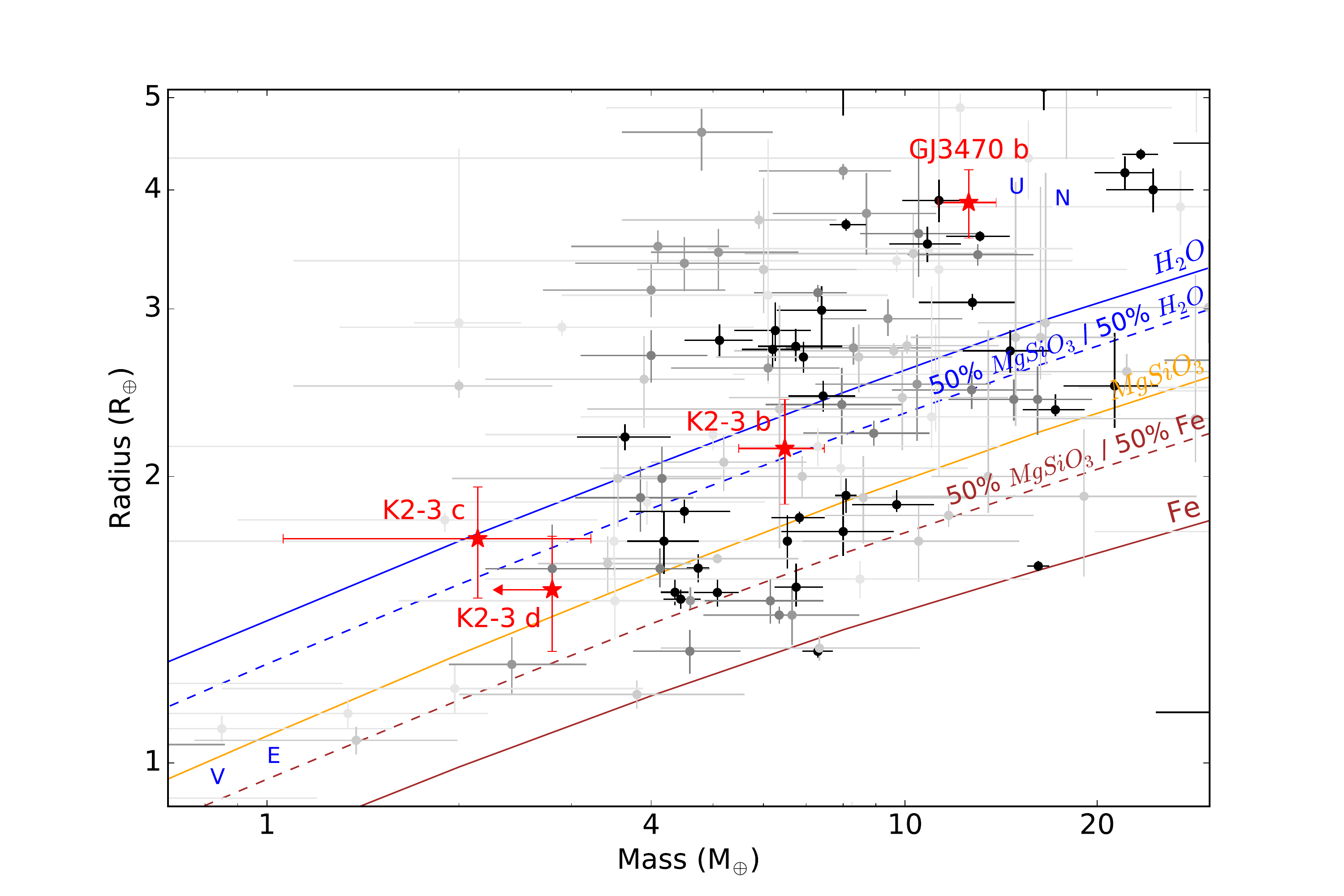}
\caption{Mass-radius diagram for planets between the size of Earth and Neptune (darker points for smaller error). The compositional curves listed are theoretical models \citep{Zeng2016} for planets with an iron (brown), silicate (orange), and water (blue) composition. K2-3 b, c, and GJ3470 b (red stars) are shown with 1$\sigma$ uncertainties along with the K2-3 d 3$\sigma$ upper limit on mass. All four planets likely have volatile-rich envelopes. 
\label{fig:mrdiagram}
}
\end{figure}

It is important to characterize potential targets to determine precise mass and surface gravity measurements, as these parameters will affect the interpretation of future transmission spectroscopy observations. We examined the potential atmospheric composition of these planets in two ways. First, we investigate their potential compositions in a mass-radius diagram (\autoref{fig:mrdiagram}). \citet{Fulton2017} describes a bimodality in occurrence rates of small planets in terms of planet radius with a gap between 1.5 and 2.0 \rearth. This distribution in radius suggests a similar distribution in planet composition, where planets smaller than 1.5 \rearth\ are super-Earths and planets 2.0--3.0 \rearth\ are sub-Neptunes. The three K2-3 planets fall in three different places relative to the radius gap \citep{Fulton2017}; planet b lies above, planet c is within the gap, and planet d is just below. 

We examine the bulk composition of these four planets in the context of other super-Earth and sub-Neptune planets (\autoref{fig:mrdiagram}).  GJ3470 b occupies the same mass-radius space as our own ice giants, Uranus and Neptune, and likely also has a substantial volatile envelope. Depending on its core composition, GJ3470 b has between 4\% and 13\% H/He \citep{Lopez2014}. K2-3 b and c both have a bulk density consistent with a mixture of silicates and water. As a water planet is an unlikely product of planet formation, they likely have iron-silicate cores with a small volatile envelope. Assuming an Earth-like core, K2-3 b and c both have about 0.5\% H/He by mass \citep{Lopez2014}. However, K2-3 c is also consistent with no volatile atmosphere given a sufficient amount of lighter material in the core, and the 3$\sigma$ mass measurement is consistent with an Earth-like composition. K2-3 d is potentially the lightest planet compared to others of similar radii; it needs substantial volatiles to explain it's placement on the mass-radius diagram. The two main interpretations are: (1) the planet is sufficiently low-mass to not detect its signal, requiring a significant volatile percentage, or (2) we have not adequately accounted for the stellar activity RV signal in this analysis, therefore, the actual mass of planet d is higher than listed here.

Our mass measurements of K2-3 b and c are within 1$\sigma$ of \citet{Almenara2015} and \citet{Dai2016}. Our mass measurement of K2-3 d is within 3$\sigma$ of \citet{Dai2016} and 4$\sigma$ of \citet{Almenara2015}. Our measurements of K2-3 b is within 1$\sigma$ of \citet{Damasso2018}, K2-3 c is within 2$\sigma$, and K2-3 d is within 2$\sigma$ of their RV fit and within 3$\sigma$ of their injection/recovery tests. 
We have improved the precision of the mass measurement of all three planets compared to previous measurements. However, due to the potential stellar activity contamination, use caution with the measurement for K2-3 d. 

\begin{figure}[h]
\hspace*{-.5cm}  
\includegraphics[width=3.5in]{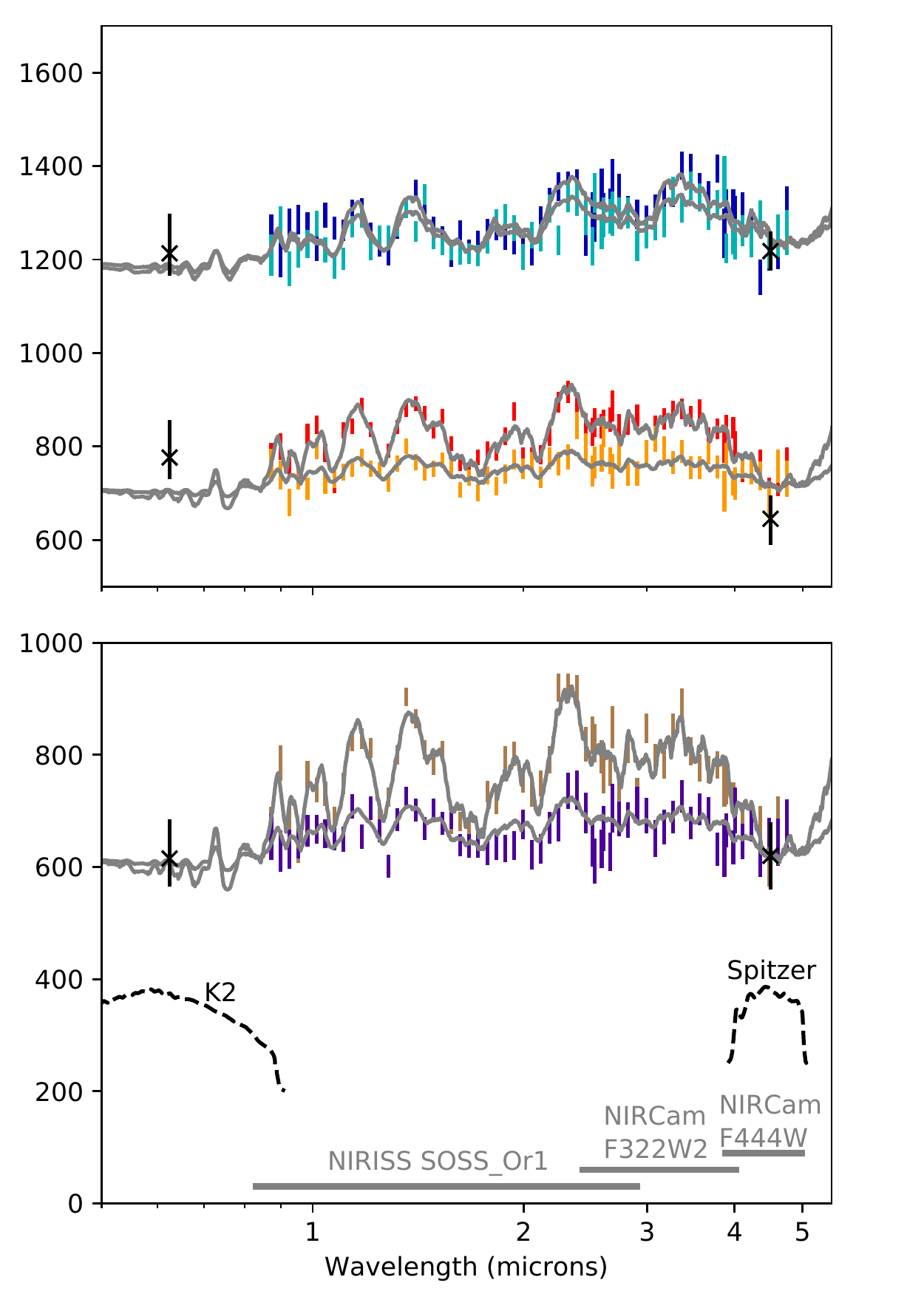}
\caption{Simulated transmission spectra (grey) of K2-3 b (blue/green, top) and c (red/orange, middle) for their 1$\sigma$ low-mass and high-mass cases and spectra for K2-3 d (brown/purple, bottom) for the upper 1$\sigma$ and 2-$\sigma$ cases. The error bars refer to simulated JWST observations with PandExo \citep{Pandexo}. \kt\ and \spitzer\ datapoints and bandpasses are shown in black. JWST instrument wavelength ranges are shown in grey. Note the break in the y-axis used for clarity.
\label{fig:transpectra}
}
\end{figure}

We then simulated model transmission spectra for the K2-3 planet system using ExoTransmit \citep{Kempton2017} to examine their possible atmospheric compositions (\autoref{fig:transpectra}). Two spectra were created for planet b and c according to the 1$\sigma$ lower and upper bounds on the mass. Two spectra were created for planet d according to the upper 2$\sigma$ and upper 3$\sigma$ mass, as the mass measurement is consistent with zero. Our assumptions include no clouds, chemical equilibrium, a 100 M/H ratio, and the 1 bar radius equals the transit radius. 

The transit depth was adjusted to match the \kt\ \citep{Crossfield2015} and \spitzer\ transit depths. Simulated JWST observations and error bars are superimposed on top of the spectra using PandExo\footnote{We present a wrapper for easier PandExo simulations, available at https://github.com/iancrossfield/jwstprep.} \citep{Greene2016,Pandexo}. We simulated one transit for each planet with three instrument modes: NIRCam F332W2, NIRCam F444W, and NIRISS SOSS\_Or1. We used the Phoenix grid models to simulate a stellar spectrum with a magnitude of 8.56 $K$ mag, temperature of 3890 K, metallicity of 0.3, and log(g) of 4.8. For each transit, we included a baseline of equal time to the transit time, zero noise floor, and resolution of R = 35. 

For K2-3 b, the absorption features would be observable for a true mass value within 1$\sigma$ of our mass measurement; the light and dark blue simulated datapoints are both inconsistent with a flat spectra. From this, K2-3 b is particularly a good target for future atmospheric study. For K2-3 c, the absorption features would be easily observable for a mass on the lower 1$\sigma$ side of our measurement, but would be much more difficult for the higher mass case. Lastly, K2-3 d would have distinguishable features as long as the mass is lower than our 2$\sigma$ upper limit. 

The \spitzer\ transit depths for K2-3 c and d are quite similar (\autoref{fig:allspitzer}) although their \kt\ transit depths are considerably offset. \citet{Beichman2016} also find similar \spitzer\ transit depths for K2-3 c and d. We were unable to create a model spectra for planet c that was consistent with both the \kt\ and \spitzer\ data to 1$\sigma$. However, this model did not include clouds, which could improve the fit of the model to the data \citep{Sing2016}. 

Transmission spectra can help to constrain a planet's mass further as the scale height depends on the planet's gravity \citep{deWit2013}. However, one must be careful as there are significant degeneracies between the effects of mass and composition for small planets \citep{Batalha2017}. With the mass of the planets constrained here through the RV method, further constraints can be put on the atmospheric composition from the transmission spectra. 

These planets are example training cases for future TESS planets. TESS will find a large sample of bright systems around nearby stars \citep{Ricker2014,Sullivan2015,ballard2018}. These types of planets will be ideal for JWST atmospheric observations due to their bright host stars. Prior to transmission spectroscopy observations, these systems will need to be followed up in a similar method as described in this paper to determine the planet masses in order to correctly interpret the spectra. 

\section{Conclusion}

In summary, we report improved masses, radii, and densities for four planets in two systems, K2-3 and GJ3470, derived from a combination of new RV, photometry, and transit observations. Our primary results are as follows.

Transit follow-ups are key for refining planet ephemerides sufficiently for future characterization. Extending the observation baseline with \spitzer\ greatly narrows the projected transit window. Our uncertainties are 20 times smaller than the original \kt\ data, which decreases the 3$\sigma$ uncertainty in the JWST era for planet d from $\sim$25 hours to under 30 minutes (\autoref{fig:spitzertime}). Our additional \spitzer\ data improve the ephemeris for the K2-3 planets to one-thirds that of \citet{Beichman2016}. See Section~\ref{sec:spitzer} for our \spitzer\ analysis and discussion.

$S_\mathrm{HK}$ may not be a good indicator for stellar activity in M dwarfs. For GJ3470, there was little to no correlation between the RVs and $S_\mathrm{HK}$; however, the rotation period found by our photometric monitoring was present in our RV data. For K2-3, although there was no correlation with $S_\mathrm{HK}$, our Evryscope photometry showed clear periodicity near the orbital period of K2-3d. Photometry and H$\alpha$ can be useful diagnostics for M-dwarf stellar rotation periods instead of $S_\mathrm{HK}$ \citep{Newton2017,Robertson2015,Damasso2018}. See Sections~\ref{sec:k2phot} and~\ref{sec:gjrot} for a description of our $S_\mathrm{HK}$ values and stellar activity discussion. 

Photometric monitoring of planet-hosting stars is important to determine the stellar rotation period and spot modulation to therefore separate the stellar activity from the planet-induced RV signals. This is especially important for planetary systems with low-amplitude RV signals as these signals may be hidden by stellar activity. We used a GP trained on our photometry to increase the accuracy of our RV fits. See Section~\ref{sec:rv} for our RV analysis including this GP. 

From our radial velocity analysis, we determined the mass of GJ3470 b to nearly ten sigma (M$_b$ = $12.58^{+1.31}_{-1.28}$ \mearth), see Section~\ref{sec:gjrv}. We additionally constrained the planet eccentricity ($e_b$ = 0.114$^{+0.52}_{-0.51}$) from our RV analysis and a measured secondary eclipse from \Spitzer. Non-zero eccentricities may be an emerging clue on how warm-Neptunes form and migrate. 

We have determined an upper limit on the mass of K2-3 d of \kdmass\ \mearth. With such a low mass, this planet is consistent with having a substantial volatile envelope which decreases its chance for habitability. As such, \ktt\ likely hosts three sub-Neptune planets instead of super-Earth planets. These planets present an interesting case for transmission spectroscopy observations of temperate sub-Neptunes. See Section~\ref{sec:disc} for simulated transmission spectra of these three planets. 

\acknowledgments
Acknowledgements:
M.R.K acknowledges support from the NSF Graduate Research Fellowship, grant No. DGE 1339067.
I.J.M.C.\ acknowledges support from NASA through  K2GO grant 80NSSC18K0308 and from NSF through grant AST-1824644. 
G.W.H. acknowledges long-term support from NASA, NSF, Tennessee State University, and the State of Tennessee through its Centers of Excellence program.
O.F. acknowledges funding support by the grant MDM-2014-0369 of the ICCUB (Unidad de Excelencia 'Mar{\'i}a de Maeztu')
This work is based [in part] on observations made with the Spitzer Space Telescope, which is operated by the Jet Propulsion Laboratory, California Institute of Technology under a contract with NASA.
This research has made use of the Exoplanet Follow-up Observing Program (ExoFOP), which is operated by the California Institute of Technology, under contract with the National Aeronautics and Space Administration.

The authors wish to recognize and acknowledge the very significant cultural role and reverence that the summit of Maunakea has always had within the indigenous Hawaiian community. We are most fortunate to have the opportunity to conduct observations from this mountain.

\facility{Keck:I, Spitzer, TNG}

\software{PandExo \citep{Pandexo}, photutils \citep{Bradley2016}, emcee \citep{emcee}, PyTransit (Parviainen 2015), batman \citep{Kreidberg2015}, RadVel \citep{Fulton2018}, ExoTransmit \citep{Kempton2017}}

\bibliography{K23format6.bib}{}
\bibliographystyle{aasjournal}

\end{document}